\documentclass[12pt]{article}
\usepackage{epsfig,latexsym}
\begin{document}
%title page stolen off Kev
\thispagestyle{empty}
\vspace*{-1 cm}
\vspace*{-1.2in}
\begin{flushright}
{SHEP 97-06} \\ April 1997 \\
\end{flushright}
 \vspace*{1.0 in}
 \begin{center}
{\Large \bf Derivative expansion of the renormalization
             group in O(N) scalar field theory} \\
\vspace*{1cm} {\bf Tim R. Morris} \\
\vspace*{.3cm}
 and \\
\vspace*{0.3cm}
{\bf Michael D. Turner} \\ \vspace*{.3cm} {\it Physics Department,
University of Southampton} \\ {\it Highfield, Southampton SO17 1BJ,
U.K.} \\
\vspace*{3 cm}
{\bf ABSTRACT} \\
\end{center}
We apply a derivative expansion to the Legendre effective action flow
equations of $O(N)$ symmetric scalar field theory, making no other
approximation. 
We calculate the critical
exponents $\eta$, $\nu$, and $\omega$ at the both the leading and
second order of the expansion, associated to the three dimensional 
Wilson-Fisher fixed points, at various values of $N$.
In addition, we show how the derivative expansion 
reproduces exactly known results, 
at special values $N=\infty,-2,-4,\cdots$.

\parskip 0.3cm
\vspace*{3cm}
\vfill\eject
\setcounter{page}{1}
\voffset -1in
\vskip2.0cm

%%%%%%%%words and defs%%%%%%%%
\def\ie{\hbox{\it i.e.}{ }}
\def\etc{\hbox{\it etc.}{ }}
\def\eg{\hbox{\it e.g.}{ }}
\def\cf{\hbox{\it cf.}{ }}
\def\viz{\hbox{\it viz.}{ }}
\def\phi{\varphi}
%%%%%%%%%%%%%%%%%%%%%%%%%%%%%%

\section{Introduction}

In contrast to other analytic approaches, many different 
non-perturbative approximations
can be envisaged within the Wilsonian renormalization group
framework~\cite{a:wk}
 which preserve the structure of the continuum limit (\viz 
renormalisability)~\cite{a:timappr}. This fact alone makes
a Wilsonian renormalization group calculational framework
potentially extremely powerful, and is surely in large part responsible
for the resurgence of interest in these methods over the
last few years. This framework holds other advantages
also, for example enabling systematic searches for 
{\sl all} (non-perturbative) continuum limits with the given field 
content and symmetries, and direct calculation of 
continuum quantities~\cite{a:timappr}-\cite{a:eqs}
%,a:timtrunc,a:revii,a:eqs}
(\ie without introducing corresponding bare quantities).
Against this background, it is surely important to
test  and understand as thoroughly as possible
the reliability and accuracy of the different possible systematic 
non-perturbative approximation schemes.

Two such schemes appear particularly natural for the Wilsonian effective
action: truncations in the operator content to a finite set
of operators; or --less severely-- an expansion and
truncation in powers of space-time derivatives where no other
approximation is made ({\it a.k.a.} ``derivative expansion''). While 
carefully chosen truncations to a finite set of operators can be very
successful in particular cases~\cite{a:t+wett,a:alford,a:aoki}, there
are reasons to expect these to have limited reliability and
accuracy in 
general non-perturbative situations~\cite{a:revii}. 
It is also difficult to see how to keep
such truncations 
systematic in practice beyond the lowest orders of the powers of derivatives
they incorporate.
In contrast, there are several reasons why one should
expect derivative expansions to be well behaved~\cite{a:revii}
at least in many situations of interest, while derivative expansions
automatically yield a sequence of systematic approximations (since there
is expansion in only one `parameter').
Additionally, since a derivative expansion in effect incorporates
at each new level of approximation a new infinite set of 
operators\footnote{in bosonic theories, corresponding to all powers of
the fields}, these expansions provide the limits of reliability
and accuracy that can anyway  be reasonably hoped for from truncations to
any finite set of operators.

The lowest order in the derivative expansion, `$O(\partial^0)$', 
corresponds to the already
well established Local Potential Approximation~\cite{a:Nicoll},
in which the effective interactions are restricted to that of a general
potential, and has proved reliable and reasonably accurate
 in a large variety
of circumstances (see the references collected in \cite{a:revii,a:timmom},
and refs.\cite{a:timtrunc,a:aoki},\cite{a:WandH}-\cite{a:largen}).
%fa:hh,a:ui,a:timhalp,a:Jordi}). 
$O(\partial^2)$ -- second order in the derivative
expansion, has been explored in \cite{a:eqs},\cite{a:stuff}-\cite{a:ball}.
The derivative expansion for the Legendre effective action with infrared
cutoff (this being the one-particle
irreducible part of the Wilson/Polchinski effective 
actions~\cite{a:timappr})
preserves more of the structure~\cite{a:timappr,a:revii} 
(\eg  preservation of quantization of anomalous scaling dimensions
after derivative expansion, for 
certain cutoffs~\cite{a:timderiv,a:timmom,a:revii},
as we review later). 
 Very encouraging accuracy has been found in the results
with this method for
the non-perturbative massless quantum field theories 
of a single scalar field, corresponding
to the two and three dimensional Ising 
model fixed points~\cite{a:timderiv,a:tim2d},
 and the infinite sequence of two dimensional multicritical fixed
points~\cite{a:tim2d}. These latter results are particularly significant
since these calculations lie well outside the practical capabilities of
all other standard approximation methods 
(including lattice methods)~\cite{a:tim2d}. Excellent results have also
been found for the massive quantum field theory defined about
three dimensional Ising model fixed point~\cite{a:eqs}.

Here we extend these second order calculations to the 
non-perturbative massless 
$O(N)$ scalar field theories in three dimensions,
corresponding to the Wilson-Fisher fixed points~\cite{a:wk}.
We remind the reader that the following values of $N$ are experimentally
realised in critical phenomena: $N=0$ corresponding to polymers;
$N=1$ corresponding to `classical' critical liquid-vapour phase transitions,
alloy order-disorder transitions, and uniaxial (Ising) ferromagnets;
$N=2$ corresponding to the He$^2$ superfluid phase transition, and to planar
ferromagnets;  $N=3$ corresponding to (Heisenberg) ferromagnets; and $N=4$
corresponding to the chiral phase transition for
two quark flavours~\cite{b:zinn,a:rob}. 

Our results for these cases, compared to the worlds best present estimates,
are fair for the lowest order, and already
rather good\footnote{with the exception of $\omega$ when $N>1$}
at second order of approximation.
% (with the exception of
%$\omega$ which does not improve  at second order when $N>1$).  
As $N$ increases beyond
%larger than 
four however, the results for $\eta$ particularly become poorer
but we find good results for $\omega$,
until as $N\to\infty$, we achieve the exact results for all 
exponents~\cite{a:WandH,b:zinn}.
We show also
that at $N=-2,-4,\cdots$, the results from derivative expansion 
coincide with the known exact 
results~\cite{a:balmin2,a:min2,a:fishmin2}.

At $N=\infty$, we give explicit analytic solutions for the fixed point
effective potential, and the eigenperturbations in the potential
and kinetic terms. We obtain analytically all the corresponding
eigenvalues, even though the perturbations to the kinetic term actually
lie outside
the exactly soluble sector in large $N$ methods~\cite{a:largen}.
At the special values $N=-2,-4,\cdots$, we obtain analytically the
known exact eigenvalues but also find intruiging hints for 
some novel fixed points.

Most importantly in our opinion,
 we continue to find (for all $N$), that the 
derivative expansion is qualitatively reliable -- in the sense that
no spurious fixed points or eigenvalues are found (in contrast
to the generic situation for finite numbers of 
operators~\cite{a:timtrunc,a:patkos}). 

The organisation of the paper is as follows. In section 2 we review 
the method as set up in \cite{a:timderiv}, and provide the slight
extensions necessary for the $N$ component case. In section 3 we develop
the $O(\partial^0)$ approximation. We remind the reader why the
only approximation that actually need be
made is to truncate the derivative expansion
of the full connected two-point function. We derive the boundary
conditions required for the fixed point and eigenperturbations.
In section 4 we develop the $O(\partial^2)$ approximation,
again develop the relevant boundary conditions, identify the one
redundant operator expected, and discuss the special behaviour at $N=1$.
Section 5 deals with the case $N=\infty$. We here
derive the exact explicit solutions already mentioned
above, and identify in passing, the line of Gaussian fixed points
that appear at $N=\infty$, and the end-point responsible for
the Bardeen-Mosher-Bander phenomenon.
We point out here the possibility to search for fixed points lieing
{\sl outside} the exactly soluble sector in large $N$ methods,
and provide physical justification for our analytic expressions
for the eigenvalues for the kinetic term perturbations,
which also lie outside the exactly soluble sector.
In section 6, we show how to recover at $N=-2$ and $N=-4$,
the known exact results for certain eigenvalues.
At the same time we find at $N=-4$, that there might exist a  
non-trivial fixed point with two  exactly soluble irrational
eigenvalues. A similar pattern is expected for the other negative
even $N$. Finally, in section 7 we present and discuss the 
numerical results. In deriving these, a
number of numerical problems had to be understood and overcome,
as described in appendix A.  Appendix B provides some necessary
beyond leading order asymptotic expressions.

\section{Review}

In this section we review, extending slightly where necessary,
the derivative expansion method developed in \cite{a:timderiv}.
We take the following as our partition function,
\begin{equation}
\exp W[J] = \int {\cal D} \phi \exp \left \{ - \frac{1}{2} \phi.C^{-1}.\phi
-S_{\Lambda_0}[\phi] +J.\phi \right \}\quad.
\label{e:partfn}
\end{equation}
We use a condensed notation, so that two-point functions are regarded
as matrices in position or momentum space, as well as any internal
index space; one-point functions as vectors; and contractions
indicated by a dot. We will work in $D$ Euclidean dimensions with a
real scalar field with $N$ components: $\phi^a$, $a=1,\cdots,N$.
$S_{\Lambda_0}$ is the full bare action for the theory, and is taken to 
have an internal $O(N)$ symmetry. $C\equiv C(q,\Lambda)$ is taken to be 
an additive cutoff for convenience\cite{a:timderiv}, satisfying 
$C \rightarrow 0$ as $q \rightarrow 0$,
and $q^2 C(q,\Lambda) \rightarrow \infty$ as $q \rightarrow \infty$.

From (\ref{e:partfn}) we have
\begin{equation}
\frac{\partial}{\partial \Lambda} W[J] = - \frac{1}{2} \left \{
\frac{\delta W}{\delta J}.\frac{\partial C^{-1}}{\partial
\Lambda}.\frac{\delta W}{\delta J} + \mathrm{tr} \left (
\frac{\partial C^{-1}}{\partial \Lambda}. \frac{\delta^2 W}{\delta J
\delta J} \right ) \right \}
\end{equation}

Re-writing this in terms of the Legendre effective action,
$\Gamma_{\Lambda}$, defined by $\Gamma_{\Lambda}[\phi^c] +
\frac{1}{2}
\phi^c.C^{-1}.\phi^c =  - W[J] +J.\phi^c$, where
the classical field $\phi^c=\delta W /
\delta J$, gives
\begin{equation}
\frac{\partial}{\partial \Lambda} \Gamma_{\Lambda}[\phi^c] = -
\frac{1}{2} \mathrm{tr} \left [ \frac{1}{C} \frac{\partial C}{\partial
\Lambda} . \left ( 1 + C.\frac{\delta^2 \Gamma_{\Lambda}}{\delta
\phi^c \delta \phi^c} \right)^{-1} \right ] \quad.
\label{e:Legrge:indices}
\end{equation}
The trace represents a sum over the internal spin indices and an
integration over momentum/position space. As in ref.\cite{a:timappr} it is
convenient to write the trace as an integral over momentum space and
factor out the $D$-dimensional solid angle. At the same time, we
%re-introduce the symmetry indices, and
drop the superscript ${}^c$ on the classical field:
\begin{equation}
\frac{\partial \Gamma_{\Lambda}[\phi]}{\partial \Lambda} =
- \frac{\Omega}{2} \mathrm{tr} \int^{\infty}_0\!\!\!\!\! dq \,
\frac{q^{D-1}}{C(q,\Lambda)}
\frac{\partial C(q,\Lambda)}{\partial \Lambda} \left \langle \left [
1 + C.\frac{\delta^2 \Gamma_{\Lambda}}{\delta \phi
\delta \phi} \right ]^{-1}\!\!\! (\mathbf{q},\mathbf{-q}) \right \rangle
\label{e:rgint}
\end{equation}
where $\Omega=2/[\Gamma(D/2)(4\pi)^{D/2}]$ is the solid angle of a
$D-1$-dimensional sphere divided by $(2 \pi)^D$, the brackets $\langle
\cdots \rangle$ represent an average over all directions of the
momentum $\mathbf{q}$, and the trace now represents only the trace
over the internal indices.

At a fixed point, the field $\phi$ scales anomalously as $\phi
\sim \Lambda^{D_{\phi}}$, where $D_{\phi} = \frac{1}{2} (D-2+\eta)$
and $\eta$ is the anomalous scaling dimension. Considering the
defining expression for $\Gamma_{\Lambda}$, $\Gamma_{\Lambda}[\phi] =
- \frac{1}{2} \phi.C^{-1}.\phi - W_{\Lambda}[J] + J_.\phi$,
dimensional analysis then shows that we require $C$ to behave as
follows,
\begin{equation}
C(q,\Lambda) \rightarrow \Lambda^{\eta -2 } \tilde{C}(q^2/\Lambda^2)
\end{equation}
for some $\tilde{C}$, if we are to express $\Gamma_{\Lambda}$ as independent
of $\Lambda$ at the fixed point. From now on we  write
$C(q,\Lambda)$ as $\Lambda^{\eta -2 }
\tilde{C}(q^2/\Lambda^2)$ and drop the tilde on the scaled $C$. 
As we are only interested in the
behaviour near or at fixed points this is a sensible definition and we
can take $\eta$ to be a constant.
This is the reason for choosing
an additive cutoff: a multiplicative cutoff  $C_{IR}$ cannot scale
homogeneously (to absorb $\Lambda^\eta$), and remain consistent 
with its normalisation in the ultraviolet ({\it viz.} 
$C_{IR}\to1$ as $q\to\infty$) \cite{a:timderiv}.
Now it is necessary (for $\Gamma_\Lambda$ to be independent 
of $\Lambda$ at fixed points) to re-write the equations in terms of
dimensionless variables,
\begin{eqnarray}
\mathbf{q} & \rightarrow & \Lambda \mathbf{q} \nonumber \\
\phi(\Lambda \mathbf{q}) & \rightarrow & \Lambda^{D-D_{\phi}} 
\phi(\mathbf{q})\quad,
\label{e:rescaling}
\end{eqnarray}
and we define $t=\log (\mu /\Lambda)$ (where $\mu$  is an 
arbitrary finite energy scale).
Finally we rescale the fields and the effective action to absorb
the factor of $\Omega/2$ in (\ref{e:rgint}),
\begin{eqnarray*}
\Gamma_t & \rightarrow & \frac{\Omega}{2 \zeta} \Gamma_t \\
\phi^a & \rightarrow & \sqrt{\frac{\Omega}{2 \zeta}} \phi^a
\end{eqnarray*}
where $\zeta$ is a numerical
normalization factor to be chosen for later convenience. Upon doing
this we get,
\begin{eqnarray}
\lefteqn{(\frac{\partial}{\partial t} +D_{\phi} \Delta_{\phi}
+\Delta_{\partial} -D) \Gamma_t[\phi] =} \label{e:rgscaled} \\ & &
-\zeta \, \mathrm{tr}\int^{\infty}_0\!\!\!\! dq \, q^{D-1} \left
(\frac{q}{C(q^2)}
\frac{\partial C(q^2)}{\partial q} + 2 - \eta \right )
 \left \langle
\left [  1 + C_.\frac{\delta^2 \Gamma_t}{\delta \phi\delta
\phi} \right ]^{-1} \!\!\!(\mathbf{q},-\mathbf{q}) \right \rangle \nonumber
\end{eqnarray}

In the above $\Delta_{\phi} = \phi.\frac{\delta}{\delta \phi}$ is the
field counting operator: it counts the number of occurrences of the
field $\phi$ in a given vertex, and arises due to the scaling of the
field in (\ref{e:rescaling}). $\Delta_{\partial}$ is the momentum
counting operator plus the dimension of space $D$ and arises through
the rescaling of the momenta in equation (\ref{e:rescaling}). It can
be represented as
\begin{equation}
\Delta_{\partial} = D + \int \frac{d^D p}{(2  \pi)^D}
\phi(\mathbf{p}) p^{\mu} \frac{\partial}{\partial p^{\mu}}
\frac{\delta}{\delta \phi(\mathbf{p})}
\end{equation}
Operating on a given vertex it counts the total number of derivatives
acting on the fields $\phi$.
Equation (\ref{e:rgscaled}) will be the starting point for all the
work from now onwards. Notice that for the first time $\eta$
explicitly appears in the equation.

We write $\Gamma_t[\phi]$ as the space-time integral of an effective
Lagrangian expanded in powers of derivatives,
\begin{equation}
\Gamma_t[\phi] = \int\!\! d^Dx\,
 \left \{ V(\phi^2,t) + \frac{1}{2} K(\phi^2,t)
\left (\partial_{\mu}
\phi^a \right)^2  + \frac{1}{2} Z(\phi^2,t) \left ( \phi^a \partial_{\mu}
\phi^a \right)^2 + \cdots \right \}
\label{e:gammalo}
\end{equation}
Each linearly independent (under integration by parts) combination of
differentiated fields will carry its own general (t-dependent)
coefficient. The global $O(N)$ symmetry forces us to choose the
coefficient functions to be functions of $\phi^2$.  We will require
that the fixed point solutions for $V$,$K$,$Z$ \etc will be
non-singular for all $\phi^2$, that perturbations about these
solutions grow no faster than a power\cite{a:timhalp,a:eqs,a:revii}, and
that $K(0) \ne 0$. It will
be seen that the imposition of power law growth results in a quantized
spectrum.

It was realized some time ago that a general cutoff function
$C(q^2)$, for example based upon an exponential function, leaves
$\eta$ undetermined and dependent upon an unphysical
parameter~\cite{a:golreparam,a:revii}. 
However, dimensional analysis indicates
that if $C(q^2)$ is chosen to be homogenous in $q^2$ then
equation~(\ref{e:rgscaled}) is invariant under a global rescaling
symmetry~\cite{a:timderiv}. 
This rescaling symmetry overdetermines the equations, so
that solutions only exist for discrete values of $\eta$. We will
follow this path and take $C(q^2)=q^{2k}$, for $k$ a positive integer.
From general arguments it can be shown that the
convergence is slower the higher the value of 
$k$~\cite{a:timappr,a:timderiv}. On
the other hand, if we wish the integrals in
equation~(\ref{e:rgscaled})  to be ultra-violet convergent we must
take $k>D/2$-1~\cite{a:timderiv}. 
Therefore we take $k$ to be the smallest integer
greater than $D/2-1$, so for $D=3$ we take $k=1$.
The rescaling symmetry is made explicit by choosing the following
(non-physical) scaling dimensions, as follows from (\ref{e:gammalo})
and the definition of $\Gamma_t$.
\begin{eqnarray}
[q^{\mu}] = 1 & [\partial^{\mu}] = 1 & [\phi^a] = k+D/2 \\ \nonumber
 [V] = D & [K] = -2(k+1) & [Z] = -2(2 k + 1) - D\quad.
\label{e:scalsym}
\end{eqnarray}
The expansion is performed by substituting equation (\ref{e:gammalo})
into equation (\ref{e:rgscaled}) and expanding the right hand side up
to a maximum number of derivatives.
 The angular average in (\ref{e:rgscaled}) can be easily
computed by translating them into invariant tensors, \eg $\langle
q^{\mu} q^{\nu} \rangle = q^2 \delta^{\mu \nu} /D$ \etc From now on we
will specialize to $D=3$.

\section{Lowest Order of Approximation}

To lowest order in the expansion we drop all the derivatives from the
right hand side of (\ref{e:rgscaled}). At the same time, a reasonable
ansatz for $\Gamma_t$ would be to keep only a general potential
$V(\phi^2,t)$ in (\ref{e:gammalo}), setting $K=1$, and $Z$ \etc to zero.
It is worthwhile to remark however, that there is no need for a
further ansatz beyond that of dropping all derivative terms from
the right hand side of (\ref{e:rgscaled}),
\ie from the full connected two-point function
 (and similarly at higher orders
of the derivative expansion): the form of $\Gamma_t$ is
then determined by consistency arguments~\cite{a:timderiv}.
Substituting (\ref{e:gammalo}) into (\ref{e:rgscaled}), 
 we see that %in this cases
the coefficient functions
$K$,$Z$, \etc are determined by linear equations given by the
vanishing of the left hand side of (\ref{e:rgscaled}). Consider the
equation for $K$,
\begin{equation}
\frac{\partial}{\partial t}K(\phi^2,t) 
+ (1+\eta) \phi^2 K'(\phi^2,t) + \eta K(\phi^2,t) = 0
\end{equation}
where $'= \frac{\partial}{\partial (\phi^2)}$. 
We see that at fixed points, where $K$ is independent of $t$, 
\begin{equation}
K(\phi^2) \propto (\phi^2)^{-\eta/(1+\eta)}\quad.
\end{equation}
If $K(\phi^2)$ is to be non-singular (at $\phi=0$) and $K(0) \ne 0$,
we must have $\eta=0$. Therefore $K(\phi^2)$ is a constant. 
Using the scaling symmetry we can set this constant to be $K=1$.
If we now consider the equation for $Z$ we get
\begin{equation}
\frac{\partial}{\partial t}Z(\phi^2,t) + 
\phi^2 Z'(\phi^2,t) + Z(\phi^2,t)=  0
\end{equation}
Hence we see that at fixed points, $Z$ will satisfy
\begin{equation}
Z \propto (\phi^2)^{-1}
\end{equation}
To obtain non-singular behaviour,  we must set the constant of
proportionality to zero and thus $Z(\phi^2) \equiv 0$. 
Considering the form of a general term in the expansion we see that a
similar conclusion must hold for all other terms. A general term, $H$,
will satisfy
\begin{equation}
\frac{\partial}{\partial t}H(\phi^2,t) +  
\phi^2 H'(\phi^2,t) +n_{\phi} H(\phi^2,t)
+ n_{\partial} H(\phi^2,t) - 3 H(\phi^2,t) =0\quad,
\end{equation}
 where $n_{\phi}$ is the number of $\phi$'s occurring in the term
multiplying $H$, divided by two and $n_{\partial}$ is the number of
derivatives occurring in the term multiplying $H$. Hence we see that
at a fixed point, we have
\begin{equation}
H \propto (\phi^2)^{- (n_{\phi} + n_{\partial} -3)}\quad.
\end{equation}
Thus for terms of higher order in the expansion, where
$n_{\phi} \ge 2$ and $ n_{\partial} > 2$, $H$ will be singular
unless the constant of proportionality is set to zero. Therefore, at
leading order, the derivative expansion must reproduce the local
potential expansion at fixed points, and thus we take
\begin{equation}
\Gamma_t[\phi] = \int\!\! d^Dx \; V(\phi^2,t)
 + \frac{1}{2} (\partial_{\mu} \phi^a)^2\quad.
\label{e:LPA}
\end{equation}

Since we discard all derivatives from
$\left [ 1 +C.\frac{\delta^2 \Gamma_t[\phi]}{\delta \phi
\delta \phi} \right ]^{-1}\!\!\!(\mathbf{q},\mathbf{-q})$,
we have that $\phi$ is {\sl effectively} a constant and the operator
 diagonal in momentum space (as explained further in the next section), 
thus it is the matrix inverse of
\begin{equation}
\delta^{ab}+C(q^2)\frac{\delta^2 \Gamma_t[\phi]}
{\delta \phi^a(\mathbf{-q}) \delta \phi^b(\mathbf{q})} = \delta^{ab}+ q^2
\left(2\delta^{ab} V' +4 \phi^a \phi^b V'' + q^2 \delta^{ab}\right)
\end{equation}
The inverse and trace in (\ref{e:rgscaled}), are now 
easily accomplished by noting that a matrix $A \delta^{a b} + B
\phi^a \phi^b$ has $N-1$ eigenvalues $A$ and one eigenvalue $A+\phi^2 B$:
%Thus (\ref{e:rgscaled}) becomes
\begin{eqnarray}
\lefteqn{\frac{\partial V(\phi^2,t)}{\partial t} + \phi^2 V'(\phi^2,t) -
3 V(\phi^2,t) =} \\ & & - 4 \zeta \int_0^\infty\!\!\!\! dq \, q^2
\frac{N-1}{1+2 V'(\phi^2,t) q^2 +q^4} + \frac{1}{1+(2V'(\phi^2,t) + 4 \phi^2
V''(\phi^2,t)) q^2 +q^4} \nonumber
\end{eqnarray}
Performing the $q$-integrals,  and
setting $\zeta$ to $1/2\pi$ yields the equation at leading order,
\begin{eqnarray}
\frac{\partial V(\phi,t)}{\partial t} +  \phi^2 V'(\phi^2,t)
- 3 V(\phi^2,t)& = & - \frac{1}{\sqrt{2 + 2 V'(\phi^2,t) + 4
\phi^2 V''(\phi^2,t)}} \nonumber \\
& & - \frac{N-1}{\sqrt{2 + 2 V'(\phi^2,t)}\quad.}
\label{e:lo}
\end{eqnarray}

\begin{figure}[ht] %h=embedded, t=top of next page, (?) p=separate page
\centering
\epsfig{figure=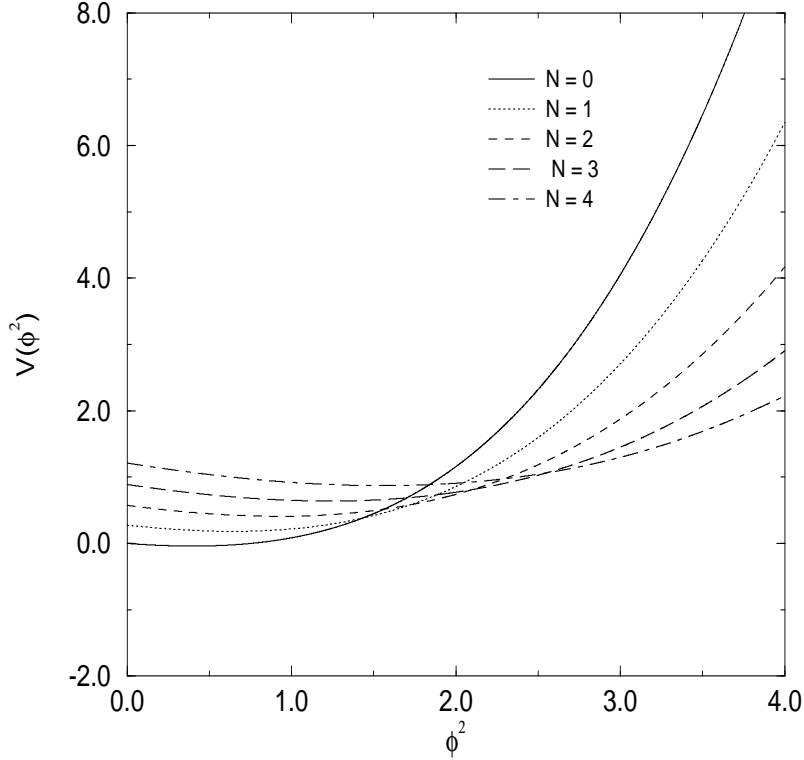,width =.8\textwidth,height=.8\textwidth}
\vskip -1cm
\caption{The potential at leading order in the derivative expansion.
%for \mbox{$D=3$}, $O(N)$ symmetric field theory, 
%for \mbox{$N=0,1,2,3,4$}.
}
\label{lo}
\end{figure}

Let us now discuss fixed point solutions of
(\ref{e:lo}), that is solutions with $\partial V/ \partial t =0$. 
At first sight it may seem that (\ref{e:lo}) has infinitely many
fixed point solutions. In fact this is not
the case, as only finitely many solutions do not end in a
singularity~\cite{a:timhalp,a:revii}. Of course this is sensible on
physical grounds, as the fixed points correspond to massless continuum
limits (\ie continuous phase transitions) with the prescribed field
content. 
%The only requirement we have imposed upon $V(\phi^2)$ so far is that
%it is non-singular for all $\phi^2$. 
Considering the form of (\ref{e:lo}) we see that either
$V(\phi^2)$ is only defined for $\phi^2 < \phi_c^2$
and ends at a singularity, at $\phi^2=\phi_c^2$ as follows,
\begin{equation}
\frac{1}{\phi_c^2} \left ( \frac{9}{16} \right )^{2/3} \left ( \phi_c^2 -
\phi^2 \right )^{4/3}
\label{e:losing}
\end{equation}
(where we have suppressed non-singular and lower order singular parts),
which we disregard as unphysical,
or it is defined for all $\phi^2>0$, and in particular for $\phi^2\to\infty$.
In this regime, 
we find that either $V(\phi^2)$ is just a 
constant (the
Gaussian fixed point), or $V(\phi^2)$ must satisfy the following,
%for large $\phi^2$,
\begin{equation}
A_v \,(\phi^2)^3 + \left(N+\frac{1}{\sqrt{5}}-1\right)
\frac{1}{4\sqrt{6A_v}\phi^2}
 +O\left((\phi^2)^{-2}\right)
\label{e:loasy}
\end{equation}
for some constant $A_v$.
If we consider linear perturbations about this solution we see that for
large $\phi^2$ the perturbation must behave as a linear combination of 
$(\phi^2)^3$ and $\exp(\frac{1}{8}[30 A_v]^{3/2} (\phi^2)^4)$. The
first perturbation merely alters the value of $A_v$, whilst we
disallow the second as it is incompatible with (\ref{e:loasy}). We see
that, apart from the Gaussian fixed point,
 the solution space of fixed point solutions defined for all
$\phi^2$, forms an isolated one parameter set,
parametrized by the value of $A_v$. Since we require
$V(\phi^2)$ also to be non-singular at the origin, 
setting $\phi^2=0$ in
equation~(\ref{e:lo}) we obtain,
\begin{equation}
-3 V(0) = - \frac{N}{\sqrt{2 + 2 V'(0)}}
\label{e:bc0lo}
\end{equation}
We now have a second order differential equation with two boundary
conditions: (\ref{e:loasy},\ref{e:bc0lo}).
 Thus we expect at most a discrete set of acceptable
solutions.  In fact we only find two:- the Gaussian fixed point,
$V(\phi^2) =
\frac{N}{3 \sqrt{2}}$,  and an
approximation to the Wilson-Fisher fixed point \cite{a:wk}.
To find this non-trivial solution we need to rely on numerical methods,
as outlined in the appendix A.  
The results of these calculations are shown in
figure \ref{lo}, for various values of $N$.

To calculate the critical exponents, we linearize about
this fixed point potential $V=V^*(\phi^2)$. Writing $V(\phi^2,t) =
V^*(\phi^2) + \delta V(\phi^2,t)$, where $\delta V(\phi^2,t)$ is given
by $ \varepsilon
\mathrm{e}^{\lambda t} v(\phi^2)$, with $\varepsilon \ll 1$, and expanding
to first order in $\varepsilon$,  we have
\begin{equation}
(\lambda -3 ) v(\phi^2) +\phi^2 v'(\phi^2)  =
\frac{(N-1) v'(\phi^2)}{\left(2+2V'(\phi^2)\right)^{3/2}}+ \frac{v'(\phi^2) + 2
\phi^2 v''(\phi^2)}{ \left(2+ 2V'(\phi^2) + 4 \phi^2 V''(\phi^2)\right)^{3/2}}
\label{e:loolin}
\end{equation}
Requiring that $v(\phi^2)$ is not singular at the origin $\phi^2=0$, we obtain
\begin{equation}
(\lambda -3 )\, v(0) =
\frac{N}{(2+2V'(0))^{3/2}}\, v'(0)\quad.
\label{e:lolinbc1}
\end{equation}
Thus for $N\ne0$ we may choose, by linearity of the perturbation, the
normalisation condition $v(0)=1$.
For $N=0$,
(\ref{e:lolinbc1}) implies that either  $\lambda=3$ or $v(0)=0$. We
discard the case of $\lambda=3$ as this corresponds to the
uninteresting case of the vacuum energy operator $v(\phi^2)\equiv1$.
Therefore for $N=0$ one 
has $v(0)=0$, and by linearity we can take $v'(0)=1$. Thus for any
$N\ge0$, we obtain two boundary conditions at the origin.
For large $\phi$, we see that the perturbation will 
be a linear combination of
\begin{equation}
(\phi^2)^{3 -\lambda}
\label{e:ppow}
\end{equation}
 and $\exp( \frac{1}{8}[30 A_v]^{3/2} (\phi^2)^4)$.
Once more we will enforce a coefficient of zero on the latter
perturbation, as we require the perturbations to grow no faster than a
power in $\phi^2$~\cite{a:timhalp,a:eqs,a:revii}.
 The imposition of this condition will ensure that
the solutions form an isolated one parameter set.
Upon also imposing the two boundary conditions at the origin,
we expect at most a discrete number of solutions to the
eigenvalue problem for $\lambda$, which is
indeed the case.  We find only one positive eigenvalue, which yields
$\nu$ through $\nu=1/\lambda$. The least negative eigenvalue yields
the first correction to scaling through $\omega=-\lambda$. The results
are summarized in the table presented in section 7.

\section{Second Order of Approximation}

The equations at the second order in the expansion are calculated by
 substituting expression (\ref{e:gammalo}) into equation
 (\ref{e:rgscaled}) and dropping all terms with more than two
 derivatives from the right hand side. Following a similar argument to
that given in the previous section,
 we see that terms with more than two derivatives are forced
 to vanish, hence at second order of the derivative expansion,
only the terms explicitly written in (\ref{e:gammalo}) survive.  
A little bit of thought shows that a similar conclusion holds at all 
higher orders
of the approximation: if we substitute an expression into the equation
that is of higher order than the order we are working at, then we are
forced to set the higher order terms to zero~\cite{a:timderiv}.

The computation of the inverse in equation
(\ref{e:rgscaled}),  is not as straightforward as in the leading
order case. We regard $[1+C.\delta^2
\Gamma_t / \delta \phi \delta \phi]^{-1}$ as a differential operator:
\begin{eqnarray}
\left [1+C.\frac{\delta^2 \Gamma_t}{\delta \phi\delta \phi}
\right ]^{-1}\!\!\!\!(\mathbf{q},\mathbf{-q}) & = & \int\! d^Dx \, d^Dy \,
\mathrm{e}^{-i \mathbf{q.x}} \left[
\delta^{ab} + C.\frac{\delta^2 \Gamma_t}
{\delta \phi^a \delta \phi^b} \right ]^{-
1}\!\!\!\!(\mathbf{x},\mathbf{y}) \, 
\mathrm{e}^{i \mathbf{q.y}} \nonumber \\
& \equiv & \int\! d^Dx \, Q\\ \mbox{where} \ \ \ \ \ \ \ \ \ \ \ \
 \ \ Q & = & \mathrm{e}^{-i \mathbf{q.x}} \left[
1 + C.\frac{\delta^2 \Gamma_t}{\delta \phi \delta \phi} \right
]^{-1}\!\!\!\mathrm{e}^{i \mathbf{q.x}}\quad.
\label{e:Q}
\end{eqnarray}
$Q^{ab}$ is a function of $\mathbf{q}$ and $\phi(\mathbf{x})$ and its
derivatives evaluated at $\mathbf{x}$, and
$C$ and $\frac{\delta^2 \Gamma_t}{\delta \phi^a \phi^b}$ are written
as differential operators:
$C(-\Box)$,
\newcommand{\fa}{\phi^a}
\newcommand{\fb}{\phi^b}
\newcommand{\p}{\partial_{\mu}}
\newcommand{\dd}{\delta^{ab}}
\newcommand{\nn}{\nonumber}
\begin{eqnarray}
\lefteqn{ \frac{\delta^2 \Gamma}{\delta \phi^a \delta \phi^b} =  
4 \fa \fb V''  + 2 \dd V'
- \dd K \Box - 2 (\Box \fa) \fb K'  - 2 \dd (\phi^c \p\phi^c) K' \p } 
\nn \\ && - 2 (\p \fa) (\p \fb) K'  - 2 (\p \fa) \fb K'
\p -4 (\p \fa) \fb (\phi^c \p \phi^c) K'' \nn \\ && + 2 \fa (\p \fb)
K' \p  + \dd (\p \phi^c)^2 K'  + 2 \fa \fb (\p \phi^c)^2 K''
 - \dd (\p \phi^c)^2 Z  \nn \\ && 
- 2 \fa (\p \fb) Z \p - 2 \fa \fb (\p \phi^c)^2 Z'
  - \dd (\phi^c \Box \phi^c) Z  - \fa
(\Box \fb) Z \nn \\ && - \fa \fb Z \Box  - 2 \fa \fb (\phi^c \Box
\phi^c) Z' - \dd (\phi^c \p \phi^c)^2 Z'  \\ && - 2 \fa (\p \fb)
(\phi^c \p \phi^c) Z'  -2 \fa \fb (\phi^c \p \phi^c) Z' \p 
- 2 \fa \fb (\phi^c \p \phi^c)^2 Z''\quad. \nn
\label{e:Gpp}
\end{eqnarray}
Define
$\nu^{ab}$ as the expression obtained by dropping all terms containing
differentials of $\phi$ from $C(q^2)
\mathrm{e}^{-i \mathbf{q.x}} \frac{\delta^2 \Gamma}
{\delta \phi^a \delta \phi^b} \mathrm{e}^{i\mathbf{q.x}}$ :
\begin{equation}
\nu^{ab}= q^2\left(4 \fa \fb V''  + 2 \dd V'+ \dd K q^2
+\fa\fb Z q^2\right) \quad.
\end{equation}
(Recall $C(q^2)=q^2$ in the present case.) Noting that 
 from (\ref{e:Q}),
\begin{equation}
Q^{ab} = \delta^{ab} - \mathrm{e}^{-i \mathbf{q.x}} C(-\Box)
\frac{\delta^2 \Gamma}{\delta
\phi^a \delta \phi^c} \left(\mathrm{e}^{i \mathbf{q.x}} Q^{cb}\right)
\label{e:Q2}
\end{equation}
we see that $Q^{ab}$ satisfies the following expression~\cite{a:timderiv},
\begin{equation}
Q^{ab} = (1 + \nu)_{ab}^{-1} + (1 + \nu)^{-1}_{ac}
\left \{
\nu^{cd}Q^{db} - \mathrm{e}^{-i \mathbf{q.x}} C(-\Box) \frac{\delta^2 \Gamma}{\delta \phi^c
\delta \phi^d} \left(\mathrm{e}^{i \mathbf{q.x}} Q^{d b}\right) \right \}
\label{e:Qit}
\end{equation}
(as may be verified by multiplying by $\delta^{ea}+\nu^{ea}$).
By construction, the term in curly brackets in (\ref{e:Qit}) is at least
of first order in the derivative expansion. 
The derivative expansion to any order may thus be derived by
 iterating expression
(\ref{e:Qit}), %to the required order, here up to two derivatives,
starting with the zeroth order $Q^{ab}=(1+ \nu)^{-1}_{ab}$. 

We  iterated just to
two derivatives [using  the fact that $1+\nu$ is of the form
$\dd+\nu^{ab}= A\dd+ B \fa \fb$, so $(1+\nu)^{-1}_{ab}=
\frac{\dd}{A}-\frac{B}{A ( A + B \phi^2)}\fa \fb$, 
and the identity
$\partial_\mu (1+\nu)^{-1}=$
$-(1+\nu)^{-1}(\partial_\mu\nu)(1+\nu)^{-1}$ ].
Even so, this is a long calculation, which we
performed using the symbolic manipulation package Form
\cite{a:Form}. Finally, in preparation for
the $N=\infty$ limit, we scale the equations as
follows~\cite{b:zinn,a:largen},
\begin{equation}
\phi  =  \sqrt{N} \tilde{\phi},\quad
V  =  N \tilde{V},\quad  K = \tilde{K},\quad Z = \tilde{Z}/N\quad.
\end{equation}
From the zeroth order part of $Q^{aa}$, we obtain
\begin{eqnarray}
\lefteqn{\frac{\partial }{\partial t}V + (1 + \eta) \phi^2 V' - 3 V =
 - \left ( 1 - \eta/4 \right ) \times} \label{e:Veqn}\\
& & \left [ \frac{1}{N\sqrt{K + \phi^2 Z}
\sqrt{2V' + 4 \phi^2 V'' +2
\sqrt{K + \phi^2 Z} } }
+ \frac{1-1/N}{\sqrt{K} \sqrt{2 V' +2\sqrt{K}} } \right ] \nonumber
\end{eqnarray}
(where here and from now on we drop the tildes on $\phi$, $V$, $K$ and $Z$).
Similarly, from the $(\partial_\mu\phi^c)^2$ part of $Q^{aa}$ we obtain,
\begin{eqnarray} 
\lefteqn{\frac{\partial }{\partial t}K+
{ (1+\eta)  \phi^2 K' + \eta K= } } \nn \\ && \frac{(\,4 - {
\eta}\,)}{\pi} \left[ {\vrule 
height 1.8em width0em depth0.80em} \right. \! \! { I}_{4, 2,
1}\, \left( \! \,{\displaystyle \frac {4}{3}}\,
{\displaystyle \frac {{ K'}^{2}\,{ \phi}}{{N}}} +
{\displaystyle \frac {4}{3}}\,{\displaystyle \frac {{ K'}\,{Z}
\,{ \phi}}{{N}}}\, \!  \right) \mbox{} + { I}_{4, 1, 2}\, \left( \! \,
{\displaystyle \frac {4}{3}}\,{\displaystyle \frac {{ K'}^{2}
\,{ \phi}}{{N}}} - 4\,{\displaystyle \frac {{ K'}\,{Z}\,{ \phi
}}{{N}}} + {\displaystyle \frac {8}{3}}\,{\displaystyle \frac {{Z
}^{2}\,{ \phi}}{{N}}}\, \!  \right) \nn  \\
 & & \mbox{} + {\displaystyle \frac {16}{3}}\,{\displaystyle
\frac {{ I}_{3, 2, 1}\,{ V''}\,{ K'}\,{ \phi
}}{{N}}} + { I}_{3, 1, 2}\, \left( \! \, - \,
{\displaystyle \frac {16}{3}}\,{\displaystyle \frac {{ V''}\,
{ K'}\,{ \phi}}{{N}}} + {\displaystyle \frac {16}{3}}\,
{\displaystyle \frac {{ V''}\,{Z}\,{ \phi}}{{N}}}\, \! 
 \right) \nn  \\
 & & \mbox{} + { I}_{2, 0, 2}\, \left( \! \,
{\displaystyle \frac {{ K'}}{{N}}} - {\displaystyle \frac {{Z}
}{{N}}} - { K'}\, \!  \right)  + { I}_{2, 2, 0}
\, \left( \! \, - \,{\displaystyle \frac {{ K'}}{{N}}} - 2\,  
{\displaystyle \frac {{ K''}\,{ \phi}}{{N}}}\, \!  \right)    
 \! \! \left. {\vrule height 1.8em width0em depth0.80em} \right]\quad, 
\label{e:Keqn}
\end{eqnarray}
where for compactness we have defined the integrals
\begin{equation}
I_{j,k,m}=\int_0^{\infty}\!\!
\frac{q^{2j}dq}{\left ( 1+ (2 V' + 4 \phi^2 V'')q^2
+(K+\phi^2 Z) q^4 \right )^k \left (1 + 2 V' q^2 + K q^4\right )^m} %\quad.
\label{e:I}
\end{equation}
These integrals can be evaluated in closed form, but (\ref{e:Keqn}) would
then be too long to display. Furthermore, as discussed in appendix A,
it proves numerically not sensible to do so. Even in this 
compact form however, the flow equation for $Z$ (which follows
from the $(\phi^c\p\phi^c)^2$ part of $Q^{aa}$) is too long to 
display~\cite{ftpsite},
\begin{equation}
\frac{\partial}{\partial t}Z + (1+\eta)\phi^2 Z' +(1+2\eta) Z  = 
\cdots 
\label{e:Zeqn}
\end{equation}

As in the previous section, we obtain the boundary conditions
for fixed point solutions, from 
the requirement that these solutions be non-singular.
Thus for
large $\phi^2$, either the solution is the trivial Gaussian fixed point
($\eta=0,V=1/3\sqrt{2},K=1,Z=0$), or  the solutions behave as follows,
\begin{eqnarray}
V(\phi^2) & \sim & \mathit{A_v} \ (\phi^2)^{\frac{3}{1+\eta}} + \cdots
\label{e:2asy1}\\ K(\phi^2) & \sim & \mathit{A_k} \
(\phi^2)^{\frac{-\eta}{1+\eta}} + \cdots \\ Z(\phi^2) & \sim &
\mathit{A_z} \ (\phi^2)^{-\frac{1+ 2 \eta}{1+ \eta}} + \cdots\quad.
\label{e:2asy2}
\end{eqnarray}
 We can also study the perturbations about these
solutions, and forcing that the perturbations grow no faster than a
power, we see that solutions
satisfying~(\ref{e:2asy1})~to~(\ref{e:2asy2}) form an isolated 
four-parameter set, that is including $\eta$. 
From the requirement that the solutions are non-singular at
the origin, substituting  $\phi^2=0$ in (\ref{e:Veqn}) -- (\ref{e:Zeqn}),
we now obtain three conditions.  Using the scaling
symmetry (\ref{e:scalsym}), we can
impose one further condition. 
We take \mbox{$K(0)=1$} (with other possible solutions
being reached by using the reparametrization invariance).
Thus with four conditions imposed on
a four-parameter set, we expect only a discrete  number of solutions,
in particular for $\eta$. In this way, 
(\ref{e:Veqn}) -- (\ref{e:Zeqn}) at fixed points, 
form {\sl non-linear} eigenvalue equations for $\eta$.
\begin{figure}[ht]
\epsfig{figure=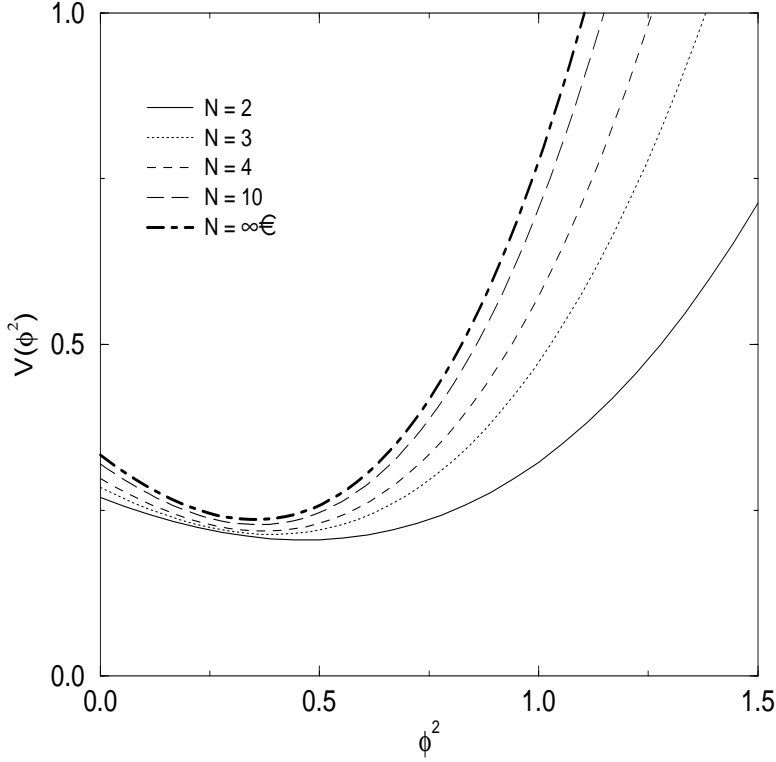,width=.8\textwidth,height=.8\textwidth}
\vskip -1cm
\caption{The potential at second order in 
the derivative expansion.} 
%for three dimensional $O(N)$ symmetric field theory, 
%for $N=2, 3, 4, 10, \infty$.}
\label{V2nd}
\end{figure}

\begin{figure}[ht]
\epsfig{figure=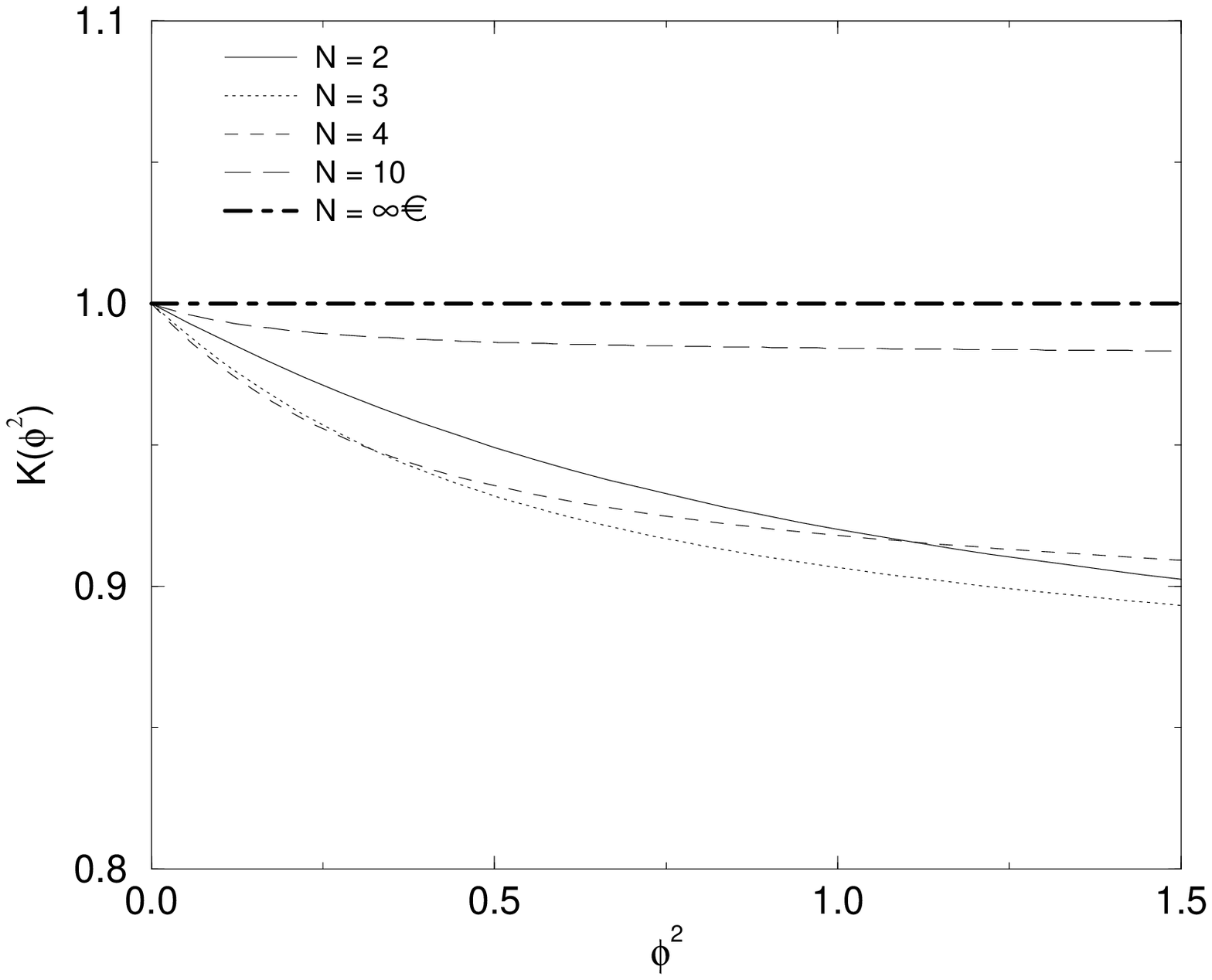,width=.8\textwidth,height=.8\textwidth}
\vskip -1cm
\caption{The $K$ component %of the Legendre effective action  
at second order in the derivative expansion. 
%for three dimensional $O(N)$ symmetric field theory, 
%for $N=2, 3, 4, 10, \infty$.
}
\label{K2nd}
\end{figure}

\begin{figure}[ht]
\vskip -2cm
\epsfig{figure=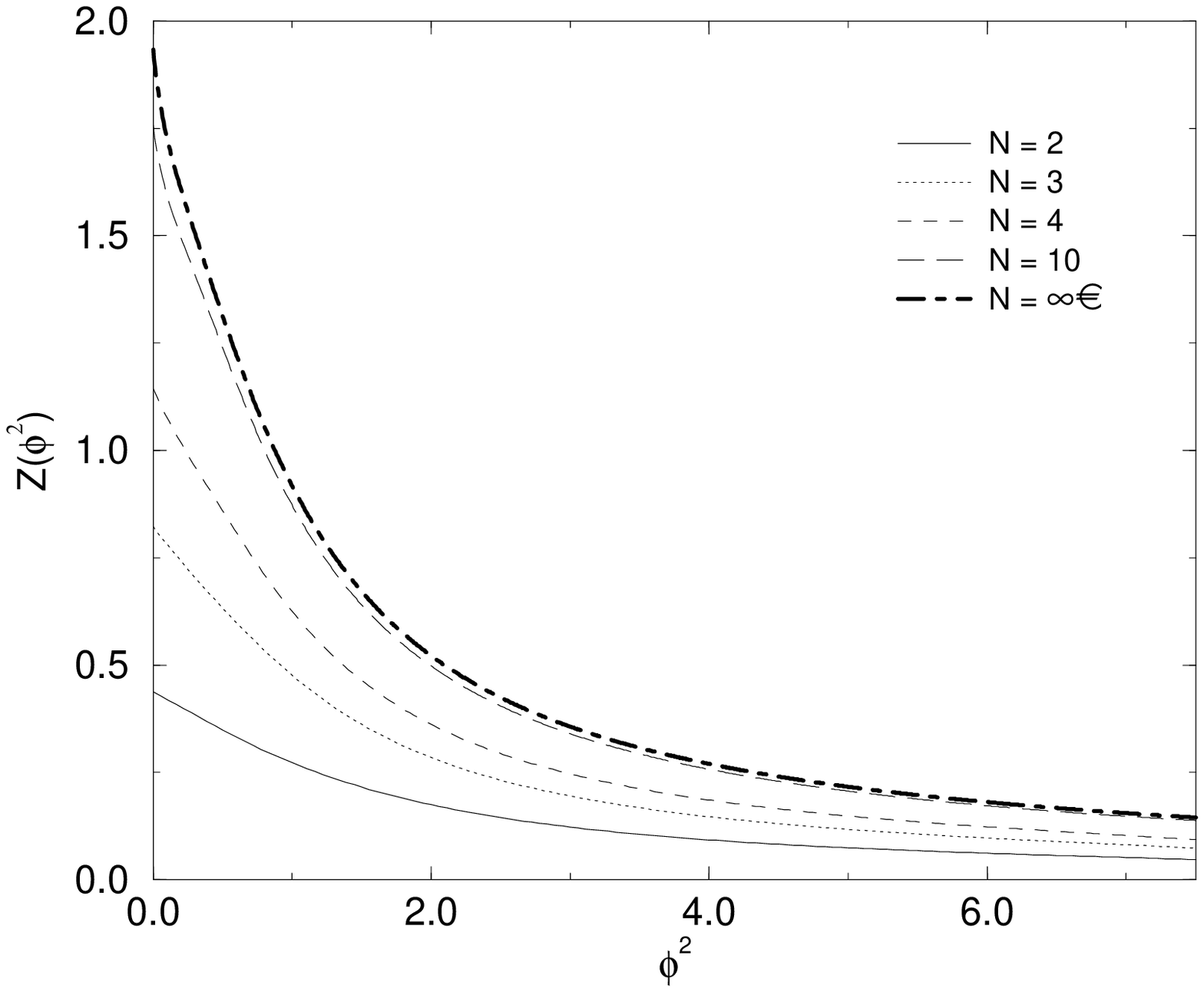,width=.8\textwidth,height=.8\textwidth}
\vskip -1cm
\caption{The $Z$ component  %of the wavefunction renormalization of the
%Legendre effective action 
at second order in the derivative expansion.
%for three dimensional $O(N)$ symmetric field theory, 
%for $N=2, 3, 4, 10, \infty$.
}
\label{Z2nd}
\end{figure}
Again we only find two solutions:- the
Gaussian point mentioned above and an approximation to the
Wilson-Fisher fixed point. The results for~$\eta$ are summarized in the
table presented in section 7,
 and the results for the fixed point solutions
shown in figures~\ref{V2nd} to~\ref{Z2nd}.

To calculate the other critical exponents (at  second order of 
approximation) we  write,
\vfill\eject

\begin{eqnarray}
V(\phi^2,t) & = & V^*(\phi^2) + \delta V(\phi^2,t) \nn \\ & = &
 V^*(\phi^2) +\varepsilon \mathrm{e}^{\lambda t} v(\phi^2) \\
 K(\phi^2,t) & = & K^*(\phi^2) + \delta K(\phi^2,t) \nn \\ & = &
 K^*(\phi^2) + \varepsilon
\mathrm{e}^{\lambda t} k(\phi^2) \\
Z(\phi^2,t) & = & Z^*(\phi^2) + \delta Z(\phi^2,t) \nn \\ & = &
Z^*(\phi^2) +\varepsilon \mathrm{e}^{\lambda t} z(\phi^2)
\end{eqnarray}
where $V^*(\phi^2)$, $K^*(\phi^2)$, $Z^*(\phi^2)$ (and $\eta^*$)
 are the fixed point
solutions calculated above, and expand the equations to first order in
$\varepsilon$.
 If we insist that $v$,$k$ and $z$ grow no faster than a
power at large $\phi^2$~\cite{a:timhalp,a:eqs,a:revii}, 
then asymptotic analysis shows that $v$,
$k$ and $z$  grow according to their scaling dimension,
\begin{eqnarray}
v(\phi^2) & \sim & \mathit{a_v} (\phi^2)^{\frac{3}{1 + \eta^*} -\lambda}
+ \cdots \\ k(\phi^2) & \sim & \mathit{a_k}
(\phi^2)^{\frac{-\eta^*}{1+\eta^*} - \lambda} + \cdots \\ z(\phi^2) & \sim
& \mathit{a_z}(\phi^2)^{-\frac{1+ 2 \eta^*}{1 + \eta^*} - \lambda} +
\cdots\quad,
\end{eqnarray}
and form an isolated four-parameter set (including $\lambda$).
As before, another three 
boundary conditions come from substituting $\phi^2=0$,
assuming non-singular limits.  Using linearity to set $v(0)=1$, we
then expect at most a discrete set of solutions, which is what is found.
%and this is what we find.  
Again we find just one positive
eigenvalue, which yields $\nu$, and determine the least negative.
These %values 
are shown in the table, presented in section 7.

It is important to recognize that there are other solutions of
the equations which do not correspond to critical
indices~\cite{a:redun}. These solutions are known as redundant
perturbations and the eigenvalue corresponding to the solution depends
on the exact form of the renormalization group chosen.
 The redundant perturbation reflects
invariances of the equations~\cite{a:redun,a:timderiv}.
In fact we expect and find only one redundant perturbation
corresponding to the reparametrization
invariance~(\ref{e:scalsym}). This has eigenvalue
\mbox{$\lambda=0$}, and (not normalised)
\begin{equation}
 v=-5\phi^2V'+ 3V,\quad\quad k=-5\phi^2K' - 4K,\quad
{\rm and}\quad z= -5 \phi^2Z' -9Z\quad.
\label{e:redun}
\end{equation}
An obvious question is why no redundant perturbations are found at the
leading order. The answer is simple: the choice of $K \equiv 1$ 
(\viz for all $t$) breaks
the reparametrization invariance, thus we should not expect to see
any redundant perturbation. Similarly we should not expect to see any
perturbation corresponding to $\phi$-translation
invariance~\cite{a:redun,a:tim2d} as the choice of $V,K$ and $Z$ being
functions of $\phi^2$ breaks this invariance.

%\subsection{The case of $N=1$}
The case of $N=1$ is special. Of course in this case
 there is only a discrete $Z_2$
internal symmetry, but also the derivative expansion (\ref{e:gammalo})
now becomes,
\begin{equation}
\Gamma_t[\phi] = \int d^Dx \left \{ V(\phi^2,t)  
+  \frac{1}{2} \left ( K(\phi^2,t) + \phi^2
Z(\phi^2,t) \right ) \left ( \partial_{\mu} \phi \right )^2 \right \}
\label{e:yabadabadoo}
\end{equation}
We see that $Z$ and $K$ no longer have a separate significance, and 
we should consider only the function \mbox{$\kappa=K + \phi^2 Z$}.
In fact at $N=1$, the $Z$ component of the fixed point solution 
to (\ref{e:Veqn})--(\ref{e:Zeqn}) is singular, diverging as $\sim 1/\phi^2$
as $\phi^2\to0$ (\cf appendix A). Taking (\ref{e:Veqn}), and forming
 (\ref{e:Keqn})$+\phi^2$(\ref{e:Zeqn}), we find that at $N=1$
all appearances 
of $K$ and $Z$ can be absorbed into $\kappa$, after
which they coincide with the
already analysed equations in ref.~\cite{a:timderiv}, as expected. 

\section{Exact Results at $N=\infty$}

It is known that in the large $N$ limit, 
the local potential approximation can effectively become
exact~\cite{a:WandH,a:aoki,a:Jordi,a:largen}, 
in the sense that (\ref{e:LPA})
then coincides with an exactly soluble subsector of the full flow equations.
In this section we will show that the derivative expansion reproduces
this result, and give an explicit
expression for the non-trivial (Wilson-Fisher)
fixed point $V^*(\phi^2)$ and for the perturbations about this fixed point. 
These solutions play an important r\^ole for us, because we obtain from
them numerically, by relaxation (\cf appendix A), the solutions at finite $N$.

Taking the $N=\infty$ limit of equations (\ref{e:Veqn}) -- (\ref{e:Zeqn}),
we obtain,
\begin{eqnarray}
\frac{\partial }{\partial t}V + (1 + \eta) \phi^2 V' - 3 V &=&
  - \frac{\left ( 1 - \eta/4 \right)}{\sqrt{K} \left (2 V'
+2\sqrt{K} \right )^{1/2}}    \label{e:Vinfin}\\
\frac{\partial}{\partial t}K + (1+\eta)\phi^2 K' +\eta K & = & -
\frac{(1-\eta/4)K'}{\sqrt{K} \left ( 2 
V' + 2\sqrt{K} \right)^{3/2}} \label{e:Kinfin}\\
\frac{\partial}{\partial t}Z + (1+\eta)\phi^2 Z' +(1+2\eta) Z & = &
\cdots \quad. \label{e:Zinfin}
\end{eqnarray} 
The equations for $K$ and
$V$ have decoupled from the equation for $Z$, and all the equations are now
\emph{first} order equations. 

The exactly soluble large $N$ limit arises when 
the {\sl interactions} in $\Gamma_t$ (or the bare action $S_{\Lambda_0}$)
are functionals only of $\phi^2$~\cite{a:largen}.
Evidently, this is true of $V$ in (\ref{e:gammalo}), while all such
derivative squared interactions can be cast (by  integration by
parts if necessary) in the form $\frac{1}{8} \left(\p\phi^2\right)^2
Z(\phi^2,t)$. This coincides with the $Z$ term in (\ref{e:gammalo})
but rules out a non-trivial $K$. Therefore we require\footnote{We are free 
to choose the normalization to be unity by the scaling symmetry 
(\ref{e:scalsym}).} $K(\phi^2,t)\equiv1$. This is indeed a solution
of (\ref{e:Kinfin}), but only if we also set $\eta=0$. 
%These are %thus the solutions that we take here.

Note that if we take $\eta=0$, then from (\ref{e:Kinfin}) the fixed
point solution has to be trivial:  $K(\phi^2,t)\equiv1$.  It
 would be very interesting to perform a thorough search of
the two parameter space (\eg parametrized by  $V(0)$ and $\eta$) of
$t$-independent solutions to (\ref{e:Vinfin},\ref{e:Kinfin}), looking
for non-singular non-trivial solutions, which would if they exist, thus
correspond to novel fixed points with non-trivial $K$ and $\eta\ne0$.
From the above analysis, such solutions for $K$ would lie outside the
exactly soluble sector of the large $N$ limit~\cite{a:largen}.

We will see that non-trivial perturbations $\delta K$ around the
Wilson-Fisher fixed point, which (from the above) also lie outside the
exactly soluble sector, yield very sensible results for the eigen-value
spectrum.

Setting $\eta=0$, and  $K(\phi^2)=1$,
 we have that the fixed
point potentials satisfy,
\begin{equation}
 \phi^2 V' - 3 V =
  - \frac{1}{ \left (2 V'
+ 2 \right )^{1/2}} 
\label{e:potninfin}
\end{equation}
Differentiating %the equation 
with respect to $\phi^2$ and setting
\mbox{$W(\phi^2) =  V'(\phi^2)$} we then have,
\begin{equation}
-2 W + \phi^2 W' = \frac{W'}{(2+2 W)^{3/2}}
\label{e:eW}
\end{equation}
This is trivially solvable as a differential equation
for $\phi^2$ with respect to $W$,
%equation can be rearranged  which is then trivially soluble:
\begin{equation}
\phi^2 = \pm A \sqrt{W} + \frac{1}{2 \sqrt{2}} \sqrt{1+W} \left ( 1+
\frac{W}{1+W} \right)
\label{e:WFW}
\end{equation}
where $A\ge0$ is a constant of integration. $A>1/\sqrt{2}$ corresponds
to the line of Gaussian fixed points that appear at 
$N=\infty$~\cite{a:Jordi,a:largen}, with the negative sign root 
for $\phi^2< 1/2\sqrt{2}$, and the positive sign root for $\phi^2>1/2\sqrt{2}$.
$A=1/\sqrt{2}$ is the end-point where the Bardeen-Mosher-Bander 
phenomenon appears~\cite{a:BMB,a:Jordi}.
$0<A<1/\sqrt{2}$ is not allowed, because there exists a range of field
values $\phi^2<\phi_c^2$ for which no real solution $W$ exists.
The value $A=0$ is the only remaining possibility, and is the solution
we take since this corresponds to the Wilson-Fisher 
fixed point~\cite{a:WandH,a:Jordi,a:largen}.
Inverting (\ref{e:WFW}) then gives two solutions, only one of which satisfies
the original equation (\ref{e:eW}) however:
\begin{equation}
W=(\phi^2)^2-1/2+\phi^2\sqrt{1+(\phi^2)^2}\quad.
\label{e:W}
\end{equation}
Integrating this solution and comparing with (\ref{e:potninfin}),
gives the potential:
\begin{equation}
V= -\frac{1}{2} \phi^2 +\frac{1}{3}(\phi^2)^3 + \frac{1}{3}
\left\{(\phi^2)^2
+1\right\}^{3/2}\quad.
\label{e:Vsol}
\end{equation}

%\newpage
Substituting $K(\phi^2)=1$ and $\eta=0$ into (\ref{e:Zinfin}), we find
\begin{eqnarray*}
%\lefteqn{\phi^2 Z'+ Z =} \\
 & & \phi^2 Z'+ Z\ = \
\frac{4-\eta}{\pi}\left\{ {\vrule height0.79em width0em depth0.79em}
\right. {  -I_{2, 0, 2}}\,{ Z'} 
 {\displaystyle +\frac {1024}{3}}\,{I_{4, 3, 2}}\,
{ V''}^{4}\,{ (\phi^2)}^{2}\\
 & & + {\displaystyle \frac {32}{3}}{ \phi^2}
  \left({\vrule height0.53em width0em depth0.53em}
 \! \,4\,{I_{6, 2, 3}} - 8\,{I_{4, 2, 3
}} + {I_{4, 2, 2}} + 4\,{I_{2, 2, 3}} - 5\,{{
I}_{2,2,2}} +   {I_{2, 2, 1}}\, \!  \right) \,{ V''}{ V'''}\\
& & + {\displaystyle \frac {64}{3}}{ \phi^2}
 \left( {\vrule height0.53em width0em depth0.53em}
 \right. \! \! 16\,{I_{5, 3, 2}}\,{Z}\,{ \phi^2} - 8\,{
I_{5, 2, 3}} - 11\,{I_{3, 2, 2}} 
 \mbox{} + 8\,{I_{3, 2, 3}} - 8\,{I_{3, 3, 2}}
 + 8\,{I_{3, 3, 1}} + 8\,{I_{5, 3, 2}} \! 
\! \left. {\vrule height0.53em width0em depth0.53em} \right) 
{ V''}^{3}\mbox{} \\
& & +  \left( {\vrule 
height0.79em width0em depth0.79em} \right. \! \! 
\mbox{} {\displaystyle \frac {256}{3}}\,{ \phi^2}\,{Z}\,
{I_{6, 3, 2}} - {\displaystyle \frac {256}{3}}\,{ \phi^2}
\,{Z}\,{I_{4, 3, 2}} + {\displaystyle \frac {256}{3}}\,{ 
\phi}^{2}\,{Z}\,{I_{4, 2, 3}}
- {\displaystyle \frac {416}{3}}\,{ \phi^2}\,{Z}\,
{I_{4, 2, 2}}  \\
 & & \mbox{} + {\displaystyle \frac {256}{3}}\,{ (\phi^2)}^{2
}\,{Z}^{2}\,{I_{6, 3, 2}}  + {\displaystyle \frac {128}{3}}
\,{I_{6, 2, 3}} - {\displaystyle \frac {256}{3}}\,{{ I
}_{4, 2, 3}}  - 64\,{I_{4, 2, 2}} + {\displaystyle \frac {
128}{3}}\,{I_{2, 2, 3}} - {\displaystyle \frac {64}{3}}\,{
I_{2, 2, 2}} \\
 & & \mbox{} - {\displaystyle \frac {64}{3}}\,{I_{2
, 2, 1}} + {\displaystyle \frac {128}{3}}\,{I_{6, 0, 5}}
  + {\displaystyle \frac {64}{3}}\,{I_{6, 3, 2}
} + {\displaystyle \frac {128}{3}}\,{I_{4, 3, 1}} - 
{\displaystyle \frac {128}{3}}\,{I_{6, 1, 4}} + 
{\displaystyle \frac {256}{3}}\,{I_{4, 1, 4}} + 
{\displaystyle \frac {80}{3}}\,{I_{4, 1, 3}} \\
 & & \mbox{}
 + {\displaystyle \frac {128}{3}}\,{I_{2, 0, 5}} + 
{\displaystyle \frac {16}{3}}\,{I_{2, 0, 3}}  - 32\,{
I_{2, 0, 4}} + {\displaystyle \frac {64}{3}}\,{I_{2
, 3, 0}} + {\displaystyle \frac {64}{3}}\,{I_{2, 3, 2}}
 - {\displaystyle \frac {128}{3}}\,{I_{2, 1, 4
}} + {\displaystyle \frac {112}{3}}\,{I_{2, 1, 3}}  \\
 & & \mbox{}  + 
{\displaystyle \frac {16}{3}}\,{I_{2, 1, 2}} - 
{\displaystyle \frac {128}{3}}\,{I_{2, 3, 1}} - 
{\displaystyle \frac {32}{3}}\,{I_{4, 0, 4}}
 - {\displaystyle \frac {256}{3}}\,{I_{4, 0, 5
}} - {\displaystyle \frac {128}{3}}\,{I_{4, 3, 2}} 
 +{\displaystyle 
\frac {256}{3}}\,{ \phi^2}\,{Z}\,{I_{4, 3, 1}} \\
& & \mbox{} - 
{\displaystyle \frac {256}{3}}\,{ \phi^2}\,{Z}\,{I_{6, 2
, 3}} 
\! \! \left. {\vrule height0.79em width0em depth0.79em} \right) 
{ V''}^{2} %\mbox{} \\
%&&
+  \left( {\vrule 
height0.79em width0em depth0.79em} \right. \! \!
  \mbox{} {\displaystyle \frac {8}{3}}\,{ \phi}^{2}\,{ Z'
}\,{I_{3, 2, 1}} + {\displaystyle \frac {8}{3}}\,{Z}\,{
I_{5, 2, 2}} - {\displaystyle \frac {40}{3}}\,{ \phi^2}
\,{ Z'}\,{I_{3, 2, 2}} - {\displaystyle \frac {64}{3}}
\,{ \phi^2}\,{ Z'}\,{I_{5, 2, 3}} \\
 & & \mbox{} - {\displaystyle \frac {40}{3}}\,{Z}\,{I_{3, 
2, 2}} + {\displaystyle \frac {8}{3}}\,{Z}\,{I_{3, 2, 1}}
 + {\displaystyle \frac {32}{3}}\,{Z}\,{I_{3, 2, 3}} - 
{\displaystyle \frac {64}{3}}\,{Z}\,{I_{5, 2, 3}} 
 + {\displaystyle \frac {32}{3}}\,{ \phi^2}\,{ 
Z'}\,{I_{7, 2, 3}} + 8\,{Z}\,{I_{3, 0, 3}} \\
 & & \mbox{} + 
{\displaystyle \frac {32}{3}}\,{Z}\,{I_{7, 2, 3}} + 
{\displaystyle \frac {32}{3}}\,{ \phi^2}\,{ Z'}\,{I_{
3, 2, 3}}   + \,{\displaystyle 
\frac {8}{3}}\,{ \phi^2}\,{ Z'}\,{I_{5, 2, 2}} 
\! \! \left . {\vrule height0.79em width0em depth0.79em}
 \right)  
 { V''} \left . {\vrule height0.79em width0em depth0.79em} \right \}\quad,
\label{e:Zifin}
\end{eqnarray*}
where the $I_{j,k,m}$ were defined in (\ref{e:I}). Needless to say, the
above repesents a great simplification 
compared to the full $Z$ equation~\cite{ftpsite}.
An analytic solution of this equation might be possible, by the methods
of ref.\cite{a:largen}, 
but the numerical solution is straightforward. Imposing the
constraint that $Z$ exists for all $\phi^2$ we get the solution shown
in figure~\ref{Z2nd}.

To calculate the critical exponents we again linearize about the fixed point
by writing $V(\phi^2,t)=V^*(\phi^2)+ \varepsilon\mathrm{e}^{\lambda t} 
v(\phi^2)$ and $K(\phi^2,t)=K^*(\phi^2)+ \varepsilon \mathrm{e}^{\lambda t}
k(\phi^2)$, where $V^*$ and $K^*$ denote the fixed point solutions
found above. From (\ref{e:Vinfin}) and (\ref{e:Kinfin}), 
%we obtain to first order in $\varepsilon$,
\begin{eqnarray}
\lambda v +\phi^2 v' -3 v &=& \frac{1}{2} \frac{k + 2
v'}{(2+2W)^{3/2}} +\frac{1}{2}\frac{k}{\sqrt{2+2W}}
 \label{e:vinfin} \\
\lambda k + \phi^2 k' & = & \frac{k'}{(2 +2W)^{3/2}}\quad, \label{e:kinfin}
\end{eqnarray}
where $W$ is given in (\ref{e:W}).
These equations are straightforwardly soluble: \eg
from (\ref{e:eW}) and  (\ref{e:kinfin}) we have by 
inspection\footnote{following an observation by Haruhiko Terao}\
\begin{equation}
k=a_k (W/2)^{-\lambda/2}\quad,
\label{e:ksol}
\end{equation}
while substituting $v=k(\phi^2)f(\phi^2)$, one finds
\begin{equation}
v(\phi^2)={\tilde a}_v \left(\frac{W}{2}\right)^{3/2-\lambda/2}
+\left\{\frac{4W^2+2W-1}{\sqrt{2+2W}} \right\}\frac{k(\phi^2)}{4}\quad.
\label{e:vsol}
\end{equation}
(These expressions may be further simplified by
defining $z=\mathrm{e}^{\sinh^{-1}\phi^2}$, then $W=z^2/2-1$ and the multiplier
of $k/4$ above, becomes $z^3-3z+1/z$.)

Noting the simple zero of $W(\phi^2)$ at $\phi^2=1/2\sqrt{2}$,
we see that (\ref{e:ksol}) and (\ref{e:vsol}) are non-singular 
if and only if: either $a_k=k=0$ and  $\lambda=3,1,-1,\cdots$, %which are 
the exactly soluble potential perturbations and spectrum
found previously~\cite{a:WandH,a:aoki,a:Jordi,a:largen};
or  ${\tilde a}_v=0$ and $\lambda=0,-2,-4,\cdots$.
The $\lambda=0$ solution 
of this set is the redundant perturbation (\ref{e:redun}).
The others lie outside the  large $N$ exactly soluble sector 
(as discussed above) and are 
thus presumably only approximate, however we see that their spectrum
can be understood, as with the potential perturbations, by assigning
scaling dimension $[\phi^2]=2$ to $\phi^2$, regarding this 
as a composite field. This interpretation
seems reasonable in view of the explicit appearance of such a field
with this dimension in traditional large $N$ approaches~\cite{b:zinn}.

\section{Exact Results at $N=-2$}

The fact that  $N=-2$ is also an exactly solvable case was
noted long ago~\cite{a:balmin2,a:min2}. The initial hope was to be
able to combine the known 
results at $N=\infty$ with those at $N=-2$, by using
Pad\'{e} approximants, to gain information about the physically
interesting cases $N=0,1,2,3$~\cite{a:balmin2,a:min2,a:fishmin2}. It
was found that when $N=-2$, the 
exponents are always Gaussian, $\eta=0$ and $\nu=1/2$.
We will show how the derivative expansion reproduces these results at
the leading order of the approximation. At the end of the section,
we also outline some interesting behaviour we observe at $N=-4$. 

Differentiating (\ref{e:lo}) with respect to $\phi^2$, and setting 
$\phi^2=0$ we have that fixed point potentials $V(\phi^2)$ satisfy,
\begin{equation}
 - 2 V'(0)=- \frac{N+2}{\{2+2V'(0)\}^{3/2}}V''(0)
\end{equation}
Thus at $N=-2$, $V'(0)=0$. Substituting this into (\ref{e:lo}), 
we find $V(0)=-\sqrt{2}/3$. Therefore in this case,
 $V(0)$ and $V'(0)$  are exactly determined, but $V''(0)$
is an {\it a priori} free parameter.  Numerically, we find 
that $V''(0)$ is fixed by the requirement that $V(\phi^2)$ be non-singular
for all $\phi^2\ge0$. We find two solutions:
the trivial tricritical Gaussian fixed point $V(\phi^2)=-\sqrt{2}/3$,
and a non-trivial Wilson-Fisher fixed point with  $V''(0) > 0$,
satisfying the required asymptotic
conditions at large $\phi^2$ (\cf equation~(\ref{e:loasy})).

Now consider the eigenperturbations. Differentiating (\ref{e:loolin})
with respect to $\phi^2$, and setting 
$\phi^2=0$ we have 
\begin{equation}
 (\lambda-2)  v'(0) = \frac{(N+2)v''(0)}{\{2 V'(0)+2\}^{3/2}}
 - 3 \frac{(N+2) v'(0) V''(0)}{\{2 V'(0)+2\}^{5/2}} \quad.
\end{equation}
Hence when $N=-2$, $\lambda=2$ or $v'(0)=0$. For the case $\lambda=2$,
we checked numerically that there exists a non-singular solution
with the required asymptotic behaviour (\ref{e:ppow}).
This eigenvalue yields $\nu=1/2$. 
For the case $v'(0)=0$, setting $\phi^2=0$ in (\ref{e:loolin}) we find
$(\lambda-3)v(0)=0$. Thus either $\lambda=3$, which corresponds to the
trivial vacuum energy perturbation $v(\phi^2)\equiv1$, or $v(0)=0$.

In the latter case we now have $v(0)=v'(0)=0$, so for
non-trivial solutions of this type we may normalise  $v''(0)=1$.
We expect from Sturm-Liouville theory~\cite{a:eqs}
that about each fixed point,
for an infinite discrete set of eigenvalues $\lambda<2$, these solutions
have the required asymptotic behaviour (\ref{e:ppow}).\footnote{This is
not at variance with \cite{a:min2} since the source term $J\cdot\phi$ in
(\ref{e:partfn}) generates `anisotropic'~\cite{a:min2} correlation functions.}
The perturbations about  the Gaussian fixed point  can be determined exactly:
 it is easy to check  from (\ref{e:loolin}), that
the eigenperturbations with the required asymptotic behaviour,
correspond to the Laguerre polynomials~\cite{a:timhalp,a:Jordi} $v(\phi^2)=
2L^{N/2-1}_n(\phi^2\sqrt{2})$
(given by the formula $L^\alpha_n(y)=\frac{1}{n!}\mathrm{e}^y y^{-\alpha}
\frac{d^n}{dy^n}\left(\mathrm{e}^{-y}y^{n+\alpha}\right)$ \cite{b:GR})
with $\lambda=3-n$, $n=0,1,\cdots$. For $n\ge2$, these indeed 
satisfy $v(0)=v'(0)=0$ and  $v''(0)=1$.

As pointed out in Fisher~\cite{a:min2} we should also expect to find
exact solutions for 
$N=-4,-6,\cdots$. For $N=-4$ we find, in a similar way to above,
that $V'''(0)$ is now an {\it a priori} free parameter, while
either $V''(0)=V'(0)=0$ and $V(0)=-2^{3/2}/3$, or 
$V''(0)=12^{3/2}/5^{5/2}$, $V'(0)=1/5$ and $V(0)=-\sqrt{20/27}$. 
We have not checked however, that non-singular
solutions actually exist with these boundary conditions (apart from
of course the Gaussian solution $V'''(0)=0$ allowed by the first set).
The first set of boundary conditions
correspond to Fisher's findings~\cite{a:min2}, since
we then obtain $(\lambda-2)(\lambda-1)v'(0)=0$,
so that the Gaussian-type eigenvalues $\lambda=1,2$ are recovered
when acceptable solutions exist with $v'(0)\ne0$. (Additionally it
may be shown that $v''(0)=0$ if $\lambda\ne1$ or 2.)
The second set leads to $(6\lambda^2-9\lambda-10)v'(0)=0$,
suggesting the existence of a non-trivial fixed point 
with two exactly soluble eigenvalues $\lambda=3/4\pm \sqrt{321}/12$.

To confirm this finding (at this level of approximation)
would require however to check that a non-singular fixed point solution
actually exists, as mentioned above, and that eigenperturbations
satisfying  (\ref{e:ppow}) for large $\phi^2$, exist with the above
eigenvalues. Equally, we expect to recover Fisher's results at
$N=-6,-8,\cdots$, and further novel candidates for fixed points.
We do not pursue these interesting questions further.

\section{Numerical Results}

\begin{table} [ht]
\renewcommand{\arraystretch}{1.5}
\hspace*{\fill}
\begin{tabular}{|c||c|c||c|c|c||c|c|c|}  \hline
$N$ &\multicolumn{2}{c||}{$\eta$} &\multicolumn{3}{c||}{$\nu$}
&\multicolumn{3}{c|}{$\omega$} \\ \hline & $O(\partial^2)$ & World &
$O(\partial^0)$ & $O(\partial^2)$ & World & $O(\partial^0)$ &
$O(\partial^2)$ & World 
\\ \hline 0 &      &.030(3) &.596&     &.590(2) &.62 &     &.81(4)
\\ \hline 1 & .054 &.035(3) &.66 &.618 &.631(2) &.63 &.897 &.80(4) 
\\ \hline 2 & .044 &.037(4) & .73 & .65 & .671(5) &.66 & .38&.79(4) 
\\ \hline 3 & .035 &.037(4) & .78 & .745 &.707(5)&.71 & .33&.78(3)
\\ \hline 4 & .022 &.025(4) & .824 & .816 &.75(1) & .75 & .42  & 
\\ \hline\hline
          10& .0054 &.025 & .94 & .95 & .88& .89 & .82 & .78 
\\ \hline 20& .0021 & .013& .96 & .98 & .94& .95 & .93 & .89 
\\ \hline 100 & .00034 &.003 & .994& .998 &.989 & .991 & .988 & .98 \\
\hline
\end{tabular}
\hspace*{\fill}
\renewcommand{\arraystretch}{1}
{\caption[Critical exponents of the 
three-dimensional Wilson-Fisher fixed point.]
{ Critical exponents of the three-dimensional Wilson-Fisher fixed point.
The first two orders of the derivative expansion are compared to
a combination of the worlds best estimates~\cite{b:zinn,a:gumph}, with 
their errors, where available. $O(\partial^0)$ is the leading order
in the derivative expansion, where $\eta$ is identically zero for all $N$,
and $O(\partial^2)$ is the second order in derivative expansion.
}}
%\label{t:2ndo}
\end{table}

In the table, 
we display the numbers only to the accuracy necessary for comparison
to the worlds best estimates. We did not find second order results
for $N=0$, for numerical reasons, as explained in appendix A.

Since, as we have seen, already 
the Local Potential Approximation yields the known exact results at
$N=-2,-4,\cdots$, and at $N=-\infty$, at the very least these 
approximations provide  physically motivated interpolations
between these exact values, which thus go beyond the early hopes
of using Pad\'e approximants~\cite{a:balmin2,a:min2,a:fishmin2}.

In fact there is quite an
impressive agreement between $O(\partial^0)$ and estimates
 by
other methods. Compared to these other methods, at first there is a gradual 
decrease in the accuracy of
the prediction for $\nu$, as $N$ increases,
whilst the prediction for $\omega$ shows a
slight improvement, until as $N\to\infty$ all exponents become exactly
determined.  The results compare
favourably with those found by other authors, both  using  a
different form of cutoff~\cite{a:t+wett,a:Jordi,a:ball,a:wet2,a:wet3}
and those obtained using a sharp cutoff~\cite{a:timtrunc,a:alford,a:aoki,a:hh}.
Ref.\cite{a:Jordi} displays a table comparing results from different forms
of the Local Potential Approximation.

At $O(\partial^2)$, for $N=1\cdots4$, the results are already rather accurate
for $\eta$ and $\nu$, showing an improvement on the leading order results. 
This is also true of $\omega$ at $N=1$, but
for $N=2,3,4$ the accuracy of  $\omega$ is poorer than at $(\partial^0)$.
For the large $N$ cases however, $\omega$
is well estimated, $\nu$ is not improved at $O(\partial^2)$,
while $\eta$ is dramatically underestimated --
eventually by about a factor of 10. 

With the results exact at $N=\infty$, we see that there is a loss of
accuracy in the {\sl approach} to this limit. Indeed 
the approximation scheme appears 'too biased'
towards the $N=\infty$ results. As we speculated
in ref.\cite{a:revii}, this may be because all
the appropriate fields have not been included in the effective action,
in this regime. Indeed,
 it is known that at large $N$, a massless bound state
field also exists at the critical point~\cite{b:zinn,a:gumph}. We should thus
expect that a derivative expansion is not so well
behaved, because the vertex
functions are hiding within them the effects of this integrated out 
massless field. To ameliorate this behaviour, we should include the
bound state explicitly as an effective $O(N)$ singlet field, then amongst the
new set of fixed points in this enlarged space will be one with the
same universal properties as the original $N$ vector model, but with
better behaved derivative expansion properties. 
We expect that similar considerations
will apply to fixed points with fermions, particularly since the bound state
fields here also correspond to the order parameter ({\it a.k.a.} fermion
condensate)~\cite{a:yuri}.

\section*{Acknowledgements}

TRM would like to thank Geoff Golner for useful comments on $N=-2$, and
PPARC for support through an advanced fellowship, and grant GR/K55738.
MDT would like to thank PPARC for support through a studentship.
%\newpage

\appendix
\section{Numerical Methods}

In this section we outline the methods used to solve numerically 
our equations. We describe in particular, how to cast the
equations manifestly as coupled second order differential
equations, how the asymptotic expressions were used to
set boundary conditions, and why it is necessary to develop
these beyond leading order for $K$ and $V$, and yet one further
order for $Z$.  We explain why relaxation was our method of
choice for these problems, and why analytic forms for the integrals
$I_{i,j,k}$ proved insufficient. We describe
 the singular behaviour seen as $N\to1$,
and the relaxation from the exact results at $N=\infty$ that
was used instead. We also point out some of the (simpler) issues
that arose on solving for the perturbations.

%We begin with the numerical methods used before discussing
%particular problems that we encountered. 

All the equations that we needed to solve numerically,
form two point boundary value problems. There are two
methods for solving such equations: shooting~\cite{b:stoer,b:nr}
and relaxation~\cite{b:nr}. Both methods were used, however
relaxation is particularly suited to equations and boundary conditions
involving complicated expressions which cannot be solved in closed form:
there is no need to write the equation  explicitly in the form
$dy/dx=f(x,y)$. It is also the best method when one needs to find
solutions that depend upon some parameter, such as the value of $N$.
Once a solution for one value of
$N$ is found, one
 can use this solution as an initial guess for a close value of
$N$. Given the nature of our problem it is not surprising that
relaxation turned out to be our principal method.

We formulate the equations as a set of non-linear, coupled
second order differential equations, as follows~\cite{a:timderiv}. 
Noting that the equation 
for $Z$,\cf (\ref{e:Zeqn}), \cite{ftpsite}, or for $N=\infty$ (\ref{e:Zifin}),
contains powers of $V'''$, we
differentiate the $V$ equation (\ref{e:Veqn})
and thus find an expression for
$V'''$ in terms of $V,V',V'',K,K',Z$ and $Z'$. 
This is substituted back into the $Z$ equation. Similarly, 
the $\phi^2=0$ boundary condition for the $Z$ equation,
contains powers of $V''(0)$.  We eliminated these in favour of
$V(0),V'(0),K(0)$ and
$Z(0)$ by using the differentiated $V$ equation.
Similar transformations are required to turn the
equations for the perturbations into manifestly
second order differential equations.

For the other set of boundary conditions we choose a value of $\phi^2$,
which we call $\phi^2_{\mathrm{asy}}$, large enough such that
the asymptotic expressions in appendix B  become a good enough
approximation to the solutions~\cite{a:timderiv,a:tim2d}.
 (These expressions go beyond leading order.
 The leading order expressions
(\ref{e:2asy1})~--~(\ref{e:2asy2}) are not sufficient, as explained
below.)  We then use the asymptotic expressions
for $V,K$ and $Z$ to provide boundary conditions for $V,V',K,K',Z$ and
$Z'$. The value of $\phi^2_{\mathrm{asy}}$ must not be chosen
so large that numerical
instability prevents us from obtaining a solution. There is a certain
amount of trial and error in deciding where to set
$\phi^2_{\mathrm{asy}}$: we cannot 
really decide where to set it until we know something about the
solution. We checked that the solutions were stable against
reasonable changes in the value of $\phi^2_{\mathrm{asy}}$.

It is necessary to develop the asymptotic expressions to
beyond the leading order in order to find a solution~\cite{a:timderiv}. 
Thus for example in (\ref{e:lo}), if we just substitute
the leading order results, \viz the first term in (\ref{e:loasy}),
for $V(\phi^2_{\mathrm{asy}})$ and 
$V'(\phi^2_{\mathrm{asy}})$, and attempt to solve (\ref{e:lo})
for $V''(\phi^2_{\mathrm{asy}})$ (as in effect takes place in the
numerical procedures),
we see that there is no real solution
for any finite $\phi^2_{\mathrm{asy}}$. Hence,
we are forced to expand the asymptotic expressions to beyond leading order. A
similar result holds for $K(\phi^2)$. However, when we consider $Z(\phi^2)$, 
we see that we need to take the asymptotic expressions to 
next-to-next-to-leading order: the correction
to the leading order asymptotic behaviour does not involve $Z''$,
hence the above procedure carried only to next-to-leading order
leaves $Z''$ incorrectly determined. In order to ensure that $Z''$ gets
determined correctly (at least to leading order), 
we need to extend the  asymptotic expression yet one further order.

In the perturbed equations it is sufficient to
use the leading order asymptotic expression~\cite{a:timderiv,a:tim2d}.
  As a check we
calculated the corrections to  the leading order behaviour and
also used these. This made no difference to the values quoted.

The solution of the
equations is complicated by the fact that they are
stiff~\cite{b:stoer,b:nr}. This arises because small perturbations
from the true solutions lead to a singular behaviour. Relaxation is
particularly adept at handling stiff problems. 
In this way, the numerical solution 
of the leading order equations proved to be
easy. It was only necessary to provide an initial guess and the
relaxation routines quickly found the solution. 

The integrals (\ref{e:I}) play an important part in the solution
of the second order equations.
It is possible to find a general analytic result for them, but
these turned out not to be useful for the present purposes.
 The integrals are
generally small numbers, but in the analytic expressions these small
numbers often arise from the 
cancellation  of large numbers. In this case roundoff errors play too 
large a part in determining their value using the analytic expression.
Even performing the calculations using higher precision Fortran
routines could not provide accuracy beyond
 four significant figures in extreme (but realised)
cases. Furthermore, whilst the 
equations (\ref{e:Veqn})~--~(\ref{e:Zeqn}) contain about three
hundred terms expressed in terms of the integrals $I_{i,j,k}$,
 particularly for the larger $i,j,k$
the algebraic expressions are so large that  the
equations expressed in closed form
stretch to between twenty five and thirty thousand
terms. The main effect of this dramatic increase is cumulative
roundoff error. The attained
level of accuracy proved totally insufficient and so numerical
methods were used.
We used an adaptive integrator from the NAG libraries. The problem
was split  into two parts: a first part from zero up to some finite
value $q_s$, and the remaining bit. To avoid roundoff errors we used an
expression arising from asymptotic analysis to calculate the second
part of the integral, and only used the numerical routines to
compute the integral over the finite range first part.
 The point $q_s$ was set to
be as small as possible (to avoid numeric round off in the numerical
integration), whilst 
still allowing the asymptotic expression to be accurate.

The numerical solution of the
second order equation is then made difficult by the large number of
integrals that need to be calculated numerically
during each iteration.  It was
soon noted that the computational power required could not be provided
by normal workstations. 
Parallel Fortran code,  making use of MPI~\cite{a:MPI}, was
developed to run on a 
sixteen node IBM SP2 system. This reduced run times for the 
relaxation code from up to one week to
 typically
between ten minutes to one hour.

Initially we tried to solve the second order equations by
relaxing away from $N=1$ using the numerical
solutions obtained in \cite{a:timderiv} as initial guesses.
However, this procedure failed
because as $N\to1$ the fixed point solution $Z(\phi^2)$
becomes rapidly sharper at the origin, diverging as $Z(\phi^2)\sim1/\phi^2$
in the limit $N=1$. This was confirmed by substituting $N=1$ 
and $K=\kappa-Z\phi^2$ [\cf  (\ref{e:yabadabadoo})]
in the $K$ equation
(\ref{e:Keqn}), using the known results~\cite{a:timderiv}
 for $\kappa$ and $V$, and solving for $Z$.
This divergence is allowed because at $N=1$ only the combination
$\kappa=K+Z\phi^2$ has any meaning and this remains well behaved,
as explained at the end of section 4. Our successful strategy 
was to first solve the simpler $N=\infty$ equations,
and then relax from $1/N=0$ to finite $N$ using these solutions
as initial guesses.

The $N=\infty$,
$V$ and $K$ equations were solved analytically as described in
section 5.
The equation (\ref{e:Zinfin})
that determines $Z$ at $N=\infty$,
was solved numerically.
Being first order, there is only one boundary
condition, which is that  the solution exists for all $\phi^2$ and thus
has the
required asymptotic behaviour. For this problem we used the shooting
method. An initial value of $Z$ was used to integrate the equation out
towards the second boundary using an eight point Runge-Kutta
integrator. In general the initial value  $Z(0)$ would be wrong: if it
was too large then the solution would tend to infinity at some finite
$\phi^2$; if it was too small the solution would tend to minus
infinity at some finite value of $\phi^2$. The correct value was then
found by  binary chop.
It was particularly difficult to get past the point
$\phi^2=1/2\sqrt{2}$. This is because at this point the right hand
side of the equations has a $0/0$ type behaviour, \ie
$Z'=n(\phi^2,Z)/d(\phi^2,Z)$, such that as $\phi^2 \rightarrow 
1/2/\sqrt{2}$, $n(\phi^2,Z) \rightarrow 0$ and $d(\phi^2,Z) \rightarrow
0$, for the true solution. There are general methods of dealing with
such singular points~\cite{b:nr}, but our equations were  sufficiently well
behaved to avoid needing to use them.

The shooting method was tried for $1/N>0$ also, 
but this proved to be  unsuitable. This
is because to use shooting effectively we need to be able to reach the
other boundary or 
have a particularly simple set of equations (\eg the $N=\infty$ equations). 
The singularity structure of the equations makes it impossible to shoot
from one side to the other and hence makes the more complicated
problem unsuited to shooting.

The main difficultly in solving the second order equation was removing
the instabilities in $Z$ near the origin. These manifested themselves
in the form of sharp spikes in $Z'$ near $\phi^2=0$. These were
removed by writing $Z(\phi^2)=s(\phi^2) \tilde{Z}(\phi^2)$, where
$s(\phi^2)$ is a known scaling function that makes $\tilde{Z}(\phi^2)$
as flat as possible. This greatly speeded up the calculations  and
removed the spikes in the solutions. However even this could not quell
the instabilities that arise as $N\to1$ where $Z(0)$ actually diverges
(as discussed above).
As a consequence we found that accurate solutions even at $N=2$ were much
harder to obtain than those of $N=3$ (say), while the singular 
behaviour at $N=1$ prevented us from relaxing to $N=0$. This is the
reason why we do not display $N=0$ results at second order in 
table 1.

In the equations for the perturbations  we calculated
 the parts dependent only upon the fixed
point solutions and stored them in a file, thus saving a large
computational overhead~\cite{a:timderiv,a:tim2d}. 
It was important that these functions, which
are the functions that 
multiplied $v,v',v'',k$ \etc in each of the linearized equations, were
well determined. The size of the $z$
equation was such that small errors in the fixed point solutions made
it difficult to determine these functions accurately, for large $\phi^2$.
This problem was solved by determining these
functions using the asymptotic expressions for $V,K$ and $Z$ (\cf appendix
B) and then matching them onto the ones calculated from the fixed point
solutions.  

To conclude, we note that we checked that the eigenvalues were stable
against reasonable changes in where $\phi^2_{\mathrm{asy}}$ was set, in
the number of mesh points, and other numerical factors.  A good indicator
of numerical accuracy was to find numerically the redundant perturbation
mentioned in section 4, and check that it matched the analytic form and
that its eigenvalue vanished to good precision.  (These properties are
only completely recovered in the limit of an infinitely fine mesh.) We
found we could numerically determine this perturbation and eigenvalue
to a high degree of accuracy for all values of $N$.

\section{Asymptotic expressions for $V,K$ and $Z$}

From the second order equations (\ref{e:Veqn})~--~(\ref{e:Zeqn}),
we determined the asymptotic behaviour of $V(\phi^2)$ and $K(\phi^2)$
for large $\phi^2$ to next-to-leading order, and $Z(\phi^2)$ to
next-to-next-to-leading order, for reasons explained in appendix A.
With $A_v$, $A_k$ and $A_z$ denoting constants, the results are
\begin{eqnarray*}
\lefteqn{V(\phi^2)\sim
{\it A_v}\,{ (\phi^2)}^{ \left( \! \,3\,\frac {1}{1 + { \eta}
}\, \!  \right) } + } \\
 & &  \left( \! \,{\displaystyle \frac {1}{24}}\,{\displaystyle 
\frac {\sqrt {6}\,(\,{N} - 1\,)\,\sqrt {1 + { \eta}}}{\sqrt {
{\it A_v}}\,\sqrt {{\it A_k}}\,{N}}} + {\displaystyle \frac {1}{24
}}\,{\displaystyle \frac {\sqrt {6}\,(\,1 + { \eta}\,)}{\sqrt {
{\it A_v}}\,{N}\,\sqrt {{\it A_k} + {\it A_z}}\,\sqrt {5 - { \eta}}
}}\, \!  \right) \,{ (\phi^2)}^{ \left( \! \,\frac { - 1 + { \eta}}{1
 + { \eta}}\, \!  \right) }
\end{eqnarray*}

\vfill
\newpage

\begin{eqnarray*}
K(\phi^2)\sim
\lefteqn{{\it A_k}\,{ (\phi^2)}^{ \left( \! \, - \,\frac {{ \eta}}{1
 + { \eta}}\, \!  \right) } + 2 \left( {\vrule 
height0.79em width0em depth0.79em} \right. \! \! {\displaystyle 
\frac {1}{864}}\sqrt {6}\,(\,1 + { \eta}\,) \left( {\vrule 
height0.43em width0em depth0.43em} \right. \! \! 4\,{ \eta}^{4}\,
{\it A_k}^{3} + 16\,{\it A_k}^{2}\,{ \eta}^{4}\,{\it A_z}} \\
 & & \mbox{} + 11\,{\it A_k}\,{ \eta}^{4}\,{\it A_z}^{2} + 48\,{ 
\eta}^{3}\,{\it A_z}^{3} + 55\,{\it A_k}\,{ \eta}^{3}\,{\it A_z}^{2}
 - 52\,{\it A_k}^{3}\,{ \eta}^{3} \\
 & & \mbox{} - 56\,{\it A_k}^{2}\,{ \eta}^{3}\,{\it A_z} - 579\,{ 
\eta}^{2}\,{\it A_k}\,{\it A_z}^{2} - 480\,{\it A_k}^{2}\,{ \eta}^{2
}\,{\it A_z} - 80\,{\it A_k}^{3}\,{ \eta}^{2} \\
 & & \mbox{} - 144\,{ \eta}^{2}\,{\it A_z}^{3} - 703\,{ \eta}\,
{\it A_k}\,{\it A_z}^{2} + 336\,{ \eta}\,{\it A_k}^{3} + 40\,{\it A_k
}^{2}\,{ \eta}\,{\it A_z} \\
 & & \mbox{} - 432\,{ \eta}\,{\it A_z}^{3} - 80\,{\it A_k}\,{\it A_z
}^{2} - 240\,{\it A_z}^{3} + 160\,{\it A_k}^{2}\,{\it A_z} \! 
\! \left. {\vrule height0.43em width0em depth0.43em} \right) 
 \left/ {\vrule height0.43em width0em depth0.43em} \right. \! \! 
 \left( {\vrule height0.44em width0em depth0.44em} \right. \! \! 
\,\sqrt {{\it A_k} + {\it A_z}} \\
 & & (\, - 4\,{\it A_k} + 2\,{\it A_k}\,{ \eta} + {\it A_z} + { \eta
}\,{\it A_z}\,)^{2}\,(\,5 - { \eta}\,)^{3/2}\,{N}\,{\it A_v}^{3/2}
\, \! \! \left. {\vrule height0.44em width0em depth0.44em}
 \right) \mbox{} + {\displaystyle \frac {1}{864}}\sqrt {6} \\
 & & \sqrt {1 + { \eta}} \left( {\vrule 
height0.43em width0em depth0.43em} \right. \! \!  \left( \! \,12
\,{ \eta}^{3}\,{N} + 28\,{ \eta}^{3} + 48\,{ \eta}\,{N} - 48\,{ 
\eta}^{2}\,{N} - 32\,{ \eta}^{2} - 48\,{ \eta}\, \!  \right) \,
{\it A_k}^{3} \\
 & & \mbox{} +  \left( {\vrule height0.43em width0em depth0.43em}
 \right. \! \! 68\,{ \eta}^{2}\,{\it A_z} - 24\,{ \eta}\,{N}\,
{\it A_z} - 12\,{ \eta}^{2}\,{N}\,{\it A_z} + 44\,{ \eta}^{3}\,
{\it A_z} + 8\,{ \eta}\,{\it A_z} \\
 & & \mbox{} + 12\,{ \eta}^{3}\,{N}\,{\it A_z} - 16\,{\it A_z} \! 
\! \left. {\vrule height0.43em width0em depth0.43em} \right) 
{\it A_k}^{2}\mbox{} +  \left( {\vrule 
height0.43em width0em depth0.43em} \right. \! \! 3\,{ \eta}^{3}\,
{N}\,{\it A_z}^{2} + 3\,{ \eta}\,{N}\,{\it A_z}^{2} + 21\,{ \eta}^{
3}\,{\it A_z}^{2} \\
 & & \mbox{} + 69\,{ \eta}\,{\it A_z}^{2} + 66\,{ \eta}^{2}\,{\it 
A_z}^{2} + 24\,{\it A_z}^{2} + 6\,{ \eta}^{2}\,{N}\,{\it A_z}^{2}
 \! \! \left. {\vrule height0.43em width0em depth0.43em} \right) 
{\it A_k}\mbox{} + 5\,{ \eta}^{3}\,{\it A_z}^{3} \\
 & & \mbox{} + 15\,{ \eta}\,{\it A_z}^{3} + 15\,{ \eta}^{2}\,{\it 
A_z}^{3} + 5\,{\it A_z}^{3} \! \! \left. {\vrule 
height0.43em width0em depth0.43em} \right)  \left/ {\vrule 
height0.43em width0em depth0.43em} \right. \! \!  \left( {\vrule 
height0.44em width0em depth0.44em} \right. \! \! \,{\it A_v}^{3/2}
\,\sqrt {{\it A_k}}\,{N} \\
 & & (\,(\,2\,{ \eta} - 4\,)\,{\it A_k} + { \eta}\,{\it A_z} + 
{\it A_z}\,)^{2}\, \! \! \left. {\vrule 
height0.44em width0em depth0.44em} \right)  \! \! \left. {\vrule 
height0.79em width0em depth0.79em} \right)   \! \,{ (\phi^2)}^{ \left(
\! \, - \,\frac {4}{1 + { \eta}}\, \! \right) } 
\end{eqnarray*}

\begin{eqnarray*}
\lefteqn{Z(\phi^2)\sim
{\it A_z}\,{ (\phi^2)}^{ \left( \! \, - \,\frac {1 + 2\,{ 
\eta}}{1 + { \eta}}\, \!  \right) } +  \left( {\vrule 
height0.93em width0em depth0.93em} \right. \! \! {\displaystyle 
\frac {1}{144}} \left( {\vrule height0.43em width0em depth0.43em}
 \right. \! \! 64\,{ \eta}^{5}\,{N} - 65\,{ \eta}^{5} + 185\,{ 
\eta}^{4} - 172\,{ \eta}^{4}\,{N} - 716\,{ \eta}^{3}\,{N}} \\
 & & \mbox{} + 652\,{ \eta}^{3} + 604\,{ \eta}^{2}\,{N} - 452\,{ 
\eta}^{2} - 2524\,{ \eta} + 2348\,{ \eta}\,{N} - 1184 \\
 & & \mbox{} + 1264\,{N} \! \! \left. {\vrule 
height0.43em width0em depth0.43em} \right) \sqrt {6} \left/ 
{\vrule height0.43em width0em depth0.43em} \right. \! \!  \left( 
\! \,\sqrt {{\it A_v}}\,{N}\,(\, - 2 + { \eta}\,)^{2}\,(\,1 + { 
\eta}\,)\,(\,5 - { \eta}\,)^{3/2}\, \!  \right)  \\
 & & \mbox{} + {\displaystyle \frac {1}{24}}\,{\displaystyle 
\frac {\sqrt {6}\, \left( \! \,20\,{ \eta}^{4} - 33\,{ \eta}^{3}
 - 6\,{ \eta}^{2} - 58\,{ \eta} - 24\, \!  \right) \,(\,{N} - 1\,
)}{(\,1 + { \eta}\,)^{3/2}\,(\, - 2 + { \eta}\,)^{2}\,{N}\,
\sqrt {{\it A_v}}}} \! \! \left. {\vrule 
height0.93em width0em depth0.93em} \right) { (\phi^2)}^{ \left( \! \,
 - \,3/2\,\frac {{ \eta} + 2}{1 + { \eta}}\, \!  \right) }
\end{eqnarray*}

\vfill
\newpage

\bibliographystyle{phaip}
\bibliography{on}

\begin{thebibliography}{10}
\bibitem{ftpsite}
The $Z$ flow equation is available as pub/9706/Z.tex,
by anonymous ftp from hep2.phys.soton.ac.uk
\end{thebibliography}
\enddocument